\documentclass[twocolumn]{aastex63}

\usepackage{booktabs}
\usepackage[T1]{fontenc}

\usepackage{graphicx}
\usepackage{amsmath}

\newcommand\doublet[1]{$\lambda\lambda$#1}	
\newcommand\OII{[O \textsc{\lowercase{II}}] }
\newcommand\OIII{[O \textsc{\lowercase{III}}] }
\newcommand\Ha{H$\alpha$ }
\newcommand\NII{[N \textsc{\lowercase{II}}] }
\newcommand\mdot{\dot{M}_{\rm cool}}
\newcommand\tcool{$t_{\rm cool}$}
\newcommand\tff{$t_{\rm ff}$}
\newcommand\msun{M$_{\odot}$}

\newcommand\msunperyear{M$_{\odot}$ yr$^{-1}$}

\received{$\rule{1cm}{0.15mm}$}
\revised{$\rule{1cm}{0.15mm}$}
\accepted{$\rule{1cm}{0.15mm}$}
\submitjournal{co-authors}

\shorttitle{Testing the Limits of AGN Feedback}
\shortauthors{Calzadilla et al.}
\graphicspath{{./}{figures/}}

\begin{document}

\title{Testing the Limits of AGN Feedback and the Onset of Thermal Instability \\
in the Most Rapidly Star Forming Brightest Cluster Galaxies} 

\correspondingauthor{Michael Calzadilla}
\email{msc92@mit.edu}

\author[0000-0002-2238-2105]{Michael S. Calzadilla}
\affil{Kavli Institute for Astrophysics and Space Research, Massachusetts Institute of Technology Cambridge, MA 02139, USA}

\author[0000-0001-5226-8349]{Michael McDonald}
\affiliation{Kavli Institute for Astrophysics and Space Research, Massachusetts Institute of Technology Cambridge, MA 02139, USA}

\author[0000-0002-2808-0853]{Megan Donahue}
\affil{Michigan State University, Physics and Astronomy Dept., East Lansing, MI 48824-2320, USA}

\author[0000-0002-2622-2627]{Brian R. McNamara}
\affil{Department of Physics and Astronomy, University of Waterloo, Waterloo, ON N2L 3G1, Canada}
\affil{Waterloo Centre for Astrophysics, University of Waterloo, Waterloo, ON N2L 3G1, Canada}
\affil{Perimeter Institute for Theoretical Physics, 31 Caroline Street North, Waterloo, ON N2L 2Y5, Canada}

\author[0000-0002-2691-2476]{Kevin Fogarty}
\affil{NASA Ames Research Center, Bldg. 245, Moffett Field, CA 94035, USA}

\author[0000-0003-2754-9258]{Massimo Gaspari}
\affil{Department of Astrophysical Sciences, Princeton University, 4 Ivy Lane, Princeton, NJ 08544, USA}

\author[0000-0002-0843-3009]{Myriam Gitti}
\affil{Dipartimento di Fisica e Astronomia (DIFA), Università di Bologna, via Gobetti 93/2, 40129 Bologna, Italy}
\affil{Istituto Nazionale di Astrofisica (INAF) – Istituto di Radioastronomia (IRA), via Gobetti 101, 40129 Bologna, Italy}

\author[0000-0001-5208-649X]{Helen R. Russell}
\affil{School of Physics, Astronomy, University of Nottingham, University Park, Nottingham NG7 2RD, UK}

\author[0000-0002-5445-5401]{Grant R. Tremblay}
\affil{Center for Astrophysics, Harvard \& Smithsonian, 60 Garden St., Cambridge, MA 02138, USA}

\author[0000-0002-3514-0383]{G. Mark Voit}
\affil{Michigan State University, Physics and Astronomy Dept., East Lansing, MI 48824-2320, USA}

\author[0000-0001-5338-4472]{Francesco Ubertosi}
\affil{Dipartimento di Fisica e Astronomia (DIFA), Università di Bologna, via Gobetti 93/2, 40129 Bologna, Italy}
\affil{INAF, Osservatorio di Astrofisica e Scienza dello Spazio, via P. Gobetti 93/3, 40129 Bologna, Italy}



\begin{abstract}
We present new, deep, narrow- and broad-band \textit{Hubble Space Telescope} observations of seven of the most star-forming brightest cluster galaxies (BCGs). 
Continuum-subtracted \OII maps reveal the detailed, complex structure of warm ($T\,{\sim}\,10^4$ K) ionized gas filaments in these BCGs, allowing us to measure spatially-resolved star formation rates (SFRs) of ${\sim}60-600$ \msunperyear.
We compare the SFRs in these systems and others from the literature to their intracluster medium (ICM) cooling rates ($\mdot$), measured from archival \textit{Chandra} X-ray data, finding a best-fit relation of ${\rm \log(SFR)} = (1.67\pm0.17) \, {\rm \log(\mdot)} + (-3.25\pm0.38)$ with an intrinsic scatter of $0.39\pm0.09$ dex.
This steeper-than-unity slope implies an increasingly efficient conversion of hot ($T\,{\sim}\,10^7$ K) gas into young stars with increasing $\mdot$, or conversely a gradual decrease in the effectiveness of AGN feedback in the strongest cool cores.
We also seek to understand the physical extent of these 
multiphase filaments that we observe in cluster cores. 
We show, for the first time, that the average extent of the multiphase gas is always smaller than the radii at which the cooling time reaches 1 Gyr, the {\tcool/\tff} profile flattens, and that X-ray cavities are observed. This implies a close connection between the multiphase filaments, the thermodynamics of the cooling core, and the dynamics of X-ray bubbles.
Interestingly, we find a one-to-one correlation between the average extent of cool multiphase filaments and the radius at which the cooling time reaches 0.5 Gyr, which may be indicative of a universal condensation timescale in cluster cores.


\end{abstract}



\section{Introduction} 
\label{sec:intro}









Most of the baryonic matter in galaxy clusters is in the form of a hot (${\sim}10^7$ K), X-ray emitting plasma that permeates the space between the galaxies, known as the intracluster medium (ICM), leaving only a few percent of the mass budget to be found in stars. In so-called ``cool core'' clusters, as particles in the ICM interact and radiate away energy, they plunge deeper into the dark matter potential well of the cluster, eventually cooling out of the hot plasma phase until becoming cold enough to form stars. However, early studies of these multiphase cooling flows \citep[e.g.][]{1994ARA&A..32..277F} found that only $1-10$\% of this cooling gas is actually observed to form stars. Apparently, most of the hot plasma in clusters is being kept hot by some heating source. This dilemma is referred to as the “cooling flow problem.” In the past decade, feedback from active galactic nuclei (AGN) has emerged as the most likely heating source energetically capable of preventing the runaway cooling of the ICM \citep[see reviews by][]{2012NJPh...14e5023M,2020NatAs...4...10G,2022PhR...973....1D}. In this scenario, the activity of an AGN is driven by the accretion of infalling material onto a supermassive black hole (SMBH), which then self-regulates its fuel supply via either radiation pressure (i.e. “quasar mode” feedback), or mechanical energy from an outburst that launches relativistic jets of plasma on tens of kpc scales (“radio mode” feedback). The tight correlation between the cooling luminosity of the ICM in relaxed clusters and the radio power \citep[e.g.][]{2021MNRAS.505.2628P} as well as the outburst power --- as measured by the work done by bubbles inflated by these relativistic jets as they expand against the ICM --- establishes the SMBH as a thermostat that adds more heat to its enviroment as the central cooling time decreases, maintaining a gentle equilibrium \citep[e.g.][]{2015ApJ...805...35H}. 

A wealth of observational evidence now backs up this AGN feedback model \citep[see review by][]{2012ARA&A..50..455F}. However, the discovery of the Phoenix cluster in 2012 has since prompted closer investigations into the limits of AGN feedback \citep{2012Natur.488..349M,2015ApJ...811..111M,2019ApJ...885...63M}. This galaxy cluster hosts the most strongly starbursting brightest cluster galaxy (BCG) and the only known possible runaway cooling flow observed in any cool core galaxy cluster. 
In the most massive clusters, with classically-inferred maximal ICM cooling rates ($\mdot = M_{\rm gas}(r < r_{\rm cool})/t_{\rm cool}$) on the order of ${>}1000$ \msunperyear, BCGs seem to be forming stars at rates of ${>}10$\% of the cooling rate \citep{2018ApJ...858...45M,2019ApJ...885...63M,2016A&A...595A.123M,2015ApJ...813..117F}. In other words, these rare, rapidly-cooling systems host AGN that appear to be incapable of the roughly two orders of magnitude suppression of cooling that we find in most other systems, perhaps signalling a gradually progressing saturation of AGN feedback \citep{2018ApJ...858...45M}. Of particular interest in these extreme cooling systems and even in more modestly star-forming clusters is the role of environment. A central ICM entropy threshold of $K_0 < 30$ keV cm$^2$, or where the cooling time is roughly $t_{\rm cool} < 1$ Gyr, has been shown to be a remarkably sensitive indicator of whether accretion onto the central SMBH fuels AGN activity or star formation ``downstream'' of the cooling flow \citep[e.g.][]{1986MNRAS.221..377N,2005ApJ...632..821P,2008ApJ...683L.107C,2008ApJ...687..899R,2017MNRAS.464.4360M,2017ApJ...851...66H,2018ApJ...853..177P}. 
Now, in addition to the standard picture of AGN feedback where heating outbursts \textit{suppress} the runaway cooling of the ICM, these same outbursts may \textit{enhance} cooling via bubbles and turbulence, and allow the feedback loop to sustain itself. 

Recent studies into how the ICM becomes thermally unstable to cooling and thus fuels the formation of stars and giant molecular clouds \citep[e.g.][]{2018ApJ...853..177P} have focused on the importance of local gravitational acceleration. In particular, the ICM was originally thought to become unstable when the ratio of the cooling time {\tcool} to the free-fall time {\tff} falls roughly below unity under the assumption of a static plane-parallel atmosphere \citep{1980MNRAS.191..399C,1986MNRAS.221..377N}. A more realistic three-dimensional atmosphere with turbulence should have a {\tcool}/{\tff} threshold of about 10 \citep[e.g.][]{2012MNRAS.419.3319M,2012MNRAS.420.3174S,2012ApJ...746...94G}. When and where this threshold is reached in the cluster atmosphere, that parcel of gas cannot persist without precipitating in a ``rain'' of cold clouds. This rain fuels feedback, which then heats up and decreases the density of the cluster core, thus increasing the cooling time ({\tcool} $\propto n_{\rm e}^{-1} k_{\rm B} T^{0.5}$) and {\tcool}/{\tff} ratio. Eventually, this halts the precipitation which turns off the feedback and allows the atmosphere to eventually cool again \citep[e.g.][]{2015ApJ...811...73L,2015ApJ...799L...1V}. This self-regulating behavior has been reproduced successfully in various simulations, producing cluster atmospheres with $10<{\rm min(t_{\rm cool}/t_{\rm ff})}<30$ \citep[e.g.][]{2012MNRAS.420.3174S,2012ApJ...746...94G,2013MNRAS.432.3401G,2014ApJ...789...54L,2015ApJ...811...73L,2017ApJ...841..133M}. Observations of cool core clusters with nebular emission, molecular gas, etc. have similarly been found to have inner {\tcool}/{\tff} ratios approaching 10, with no examples significantly below this value (with the exception of the Phoenix cluster), suggesting that this value is not a threshold but rather a floor \citep[e.g.][]{2015Natur.519..203V,2017ApJ...851...66H,2018ApJ...853..177P}.

A closely related framework for instability in the ICM is that of ``chaotic cold-accretion'' (CCA), where merger or AGN-driven turbulence induces enhanced cooling \citep{2013MNRAS.432.3401G,2015A&A...579A..62G,2017MNRAS.466..677G,2018ApJ...854..167G,2017MNRAS.471.1531P,2020MNRAS.498.4983W}. This turbulence seeds a population of cool clouds with a broad angular momentum distribution. Clouds comprising the low-end tail of this distribution fall inwards towards the center of the cluster and fuel enhanced accretion onto the SMBH, relative to the approximately Bondi accretion found in more homogeneous media \citep{2005ApJ...632..821P,2013MNRAS.432.3401G,1993MNRAS.263..323T,1995MNRAS.276..663B,2015ApJ...811..108P}. In this CCA model, the condition for thermally unstable cooling to occur is that the cooling time is small compared to the local dynamical or turbulent ``eddy'' time ({\tcool}/$t_{\rm eddy} \lesssim 1$; e.g. \citealt{2018ApJ...854..167G}), rather than an order of magnitude longer than the local freefall time (i.e. {\tcool}/{\tff} $\lesssim 10$). 
In the ``stimulated feedback'' model \citep[e.g.][]{2016ApJ...830...79M}, thermally unstable cooling happens in-situ when warm (${\sim}1$ keV) X-ray gas is uplifted in the wake of AGN-inflated radio bubbles as they rise buoyantly out of the central cluster potential, which increases the infall time and promotes the condensation of cold gas (see also \citealt{2010MNRAS.406.2023P}). Evidence for this phenomenon has been provided in a number of systems where large reservoirs (${\sim} 10^{10}$ \msun) of cold ($10-100$ K) molecular gas, observed with the \textit{Atacama Large Millimeter Array} (ALMA), is projected behind or draped around the location of X-ray cavities as seen by \textit{Chandra} (e.g. 
A1835: \citealt{2014ApJ...785...44M}, 
Phoenix: \citealt{2017ApJ...836..130R}, 
MACS 1931.8-2634: \citealt{2021A&A...649A..23C},
A1664: \citealt{2019ApJ...875...65C}, 
PKS 0745-191: \citealt{2016MNRAS.458.3134R}, 
A2597: \citealt{2018ApJ...865...13T}, 
A1795: \citealt{2017MNRAS.472.4024R}, 
2A 0335+096: \citealt{2016ApJ...832..148V}).
On the other hand, several systems show cold gas in hot halos, even without a direct correlation with bubbles
\citep[e.g.][]{2018ApJ...858...17T,2019A&A...631A..22O,2019MNRAS.489..349R,2021A&A...656A..45M,2021MNRAS.503.5179N,2021NatAs.tmp..256M}.

To distinguish between the various models that explain the onset of nonlinear thermal instabilities in the ICM, one can look at \textit{where} those thermal instabilities develop. For instance, stimulated feedback theory predicts that one ought to see condensation only up to a maximum altitude where radio bubbles could uplift lower entropy and lower altitude material, or compress higher-altitude material at the bubble's leading edge. Meanwhile, precipitation theory predicts that this condensation of cold gas should happen where {\tcool}/{\tff} $\lesssim 10$, or equivalently where the ICM entropy profile changes its slope due to heating from the AGN. One of the most successful ways to identify where multiphase cooling occurs has been via narrow-band surveys for \Ha emission in low-redshift ($z\lesssim 0.1$) clusters \citep[e.g.][]{1989ApJ...338...48H,1999MNRAS.306..857C,2007MNRAS.380...33H,2010ApJ...721.1262M,2016MNRAS.460.1758H}. As we have discussed that the ICM becomes multiphase when the central cooling time falls below ${\sim}1$ Gyr \citep[e.g.][]{2008ApJ...683L.107C}, and that the cooling time of the ICM where it is coincident with the presence of \Ha emission is ${\sim} 4 \times$ shorter than the surrounding ICM \citep[e.g.][]{2010ApJ...721.1262M}, we can use the morphology of extended \Ha nebulae to map out where the ICM is undergoing multiphase condensation. 

While \Ha emission traces warm (${\sim} 10^4$ K) gas that has already cooled but not yet formed stars or accreted onto the central SMBH, few observatories are capable of probing the redshifted \Ha wavelengths of more distant clusters ($z \gtrsim0.3$), especially with the required angular resolution to study these thin ($\lesssim1$ kpc) filamentary nebulae in detail. To address this gap in the literature, we present in this study ${\sim} 40$ orbits of new \textit{Hubble Space Telescope} (HST) data on seven of the most extremely star-forming (${\gtrsim} 100$ \msunperyear) BCGs known \citep[see e.g.][]{2018ApJ...858...45M}. Given the relatively high average redshift of these systems ($\langle z_{\rm avg}\rangle = 0.38$), we obtained deep, high-angular resolution maps of \OII\doublet{3726,3729} emission rather than H$\alpha$, 
providing a cleaner and higher signal-to-noise view of the thermally unstable regions of these extreme BCGs. 
Both \Ha and \OII have been used extensively as star formation indicators, and in this study we will assume that both probe regions of active star formation. 
Our new \OII maps will allow us to correlate line emitting gas to the cooling ICM, star forming clumps, and radio jets/bubbles. In \autoref{sec:data} we describe our HST observations, observing strategy, and data reduction in detail, as well as the archival \textit{Chandra} data that we use to investigate any correspondences between the warm (${\sim}10^4$ K) ionized \OII gas and the spatial and spectral properties of the hot (${\sim}10^7$ K) ICM. The resulting maps and star formation rates (SFRs) associated with these new observations are presented in \autoref{sec:results}, with their implications for the limits of AGN feedback and the onset of thermal instabilities discussed in \autoref{sec:discussion}. Finally, we conclude with the take-away points in \autoref{sec:conclusion}. Throughout this paper, we assume a flat $\Lambda$CDM cosmology with $H_0 = 70$ km s$^{-1}$ Mpc$^{-1}$, $\Omega_m = 0.3$, and $\Omega_{\Lambda} = 0.7$. All measurement errors are $1\sigma$ unless noted otherwise.

\section{Observations \& Data Reduction} \label{sec:data}

\subsection{Sample Selection}
Our ``extreme cooling sample'' originally consisted of six of the most star-forming (SFR$\,{\gtrsim} 100$ \msunperyear) and strongly-cooling ($\mdot\, {\gtrsim} 1000$ \msunperyear) BCGs studied in \citet{2018ApJ...858...45M}. These systems include the Phoenix cluster ($z{=}0.596$), H1821+643 ($z{=}0.299$, hereafter ``H1821''), IRAS 09104+4109 ($z{=}0.442$, ``IRAS09104''), Abell 1835 ($z{=}0.252$, ``A1835''), RX J1532.9+3021 ($z{=}0.363$, ``RXJ1532''), and MACS 1931.8-2634 ($z{=}0.352$, ``MACS1931''). SFRs for these systems, presented in \citet{2018ApJ...858...45M}, were aggregated from multiple literature sources that utilize various SFR measures (e.g. $H\alpha$, UV, or IR flux). With the exception of RXJ1532 and MACS1931, with $16+$ band HST imaging from CLASH\footnote{\url{https://archive.stsci.edu/prepds/clash/}} allowing careful stellar population modeling \citep{2017ApJ...846..103F}, AGN contamination was thought to be affecting the SFR measurements of most of these systems, making the quoted SFRs less secure, even after spatial or spectral decomposition attempts. The new observations we present below will provide more secure SFRs for these systems. In addition, RBS797 ($z{=}0.354$), which was previously estimated to have a low SFR relative to its ICM cooling rate (${\mdot}\,{\sim} 1000$ \msunperyear), was also added to our sample upon measuring a much higher updated SFR for it in a new, separate HST narrowband program, as we will show below in \autoref{subsec:SFRs}.

\begin{table}[t]
  \centering
  \caption{Summary of HST ACS/WFC Observations}
    \begin{tabular}{lrlc}
    \toprule
    \toprule
    \multicolumn{1}{l}{Name} & \multicolumn{1}{r}{t (s)} & Filters & \multicolumn{1}{c}{ID} \\
    \midrule
    \multicolumn{1}{l}{Phoenix} & 22,696 & FR601N @ 5952.431\AA & 15315 \\
    ($z=0.596$) & 5,042  & F775W & 15315 \\
          & 5,042  & F475W & 15315 \\
    \midrule
    \multicolumn{1}{l}{H1821+643} & 11,658 & FR505N @ 4835.109\AA & 15661 \\
    ($z=0.299$) & 5,512  & F850LP & 15661 \\
          & 11,658 & F550M & 15661 \\
    \midrule
    \multicolumn{1}{l}{IRAS 09104+4109} & 10,851 & FR551N @ 5375.890\AA & 15661 \\
    ($z=0.442$) & 5,166  & F625W & 15661 \\
          & 5,166  & F435W & 15661 \\
    \midrule
    \multicolumn{1}{l}{Abell 1835} & 10,582 & FR462N @ 4670.725\AA & 15661 \\
    ($z=0.252$) & 14,084 & F850LP & 15661 \\
          & 4,860  & F555W & 15661 \\
    \midrule
    \multicolumn{1}{l}{RX J1532.9+3021} & 10,454 & FR505N @ 5079.983\AA & 16494 \\
    ($z=0.363$) & 2,045  & F775W & 12454 \\
          & 2,050  & F435W & 12454 \\
    \midrule
    \multicolumn{1}{l}{MACS 1931.8-2634} & 7586  & FR505N @ 5039.147\AA & 15661 \\
    ($z=0.352$) & 2,001  & F775W & 12456 \\
          & 2,015  & F435W & 12456 \\
    \midrule
    \multicolumn{1}{l}{RBS 797} & 15,169 & FR505N @ 5046.117\AA & 16001 \\
    ($z=0.354$) & 1,200  & F814W & 10875 \\
          & 5,728  & F435W & 16001 \\
    \bottomrule
    \end{tabular}%
  \label{tab:obs_HST}%
\end{table}%

\subsection{Optical/HST}




We obtained new HST optical data for our sample during programs GO15661 and GO16494 (PI: McDonald), and GO16001 (PI: Gitti). These included narrowband data using the ramp filters on the Advanced Camera for Surveys (ACS)/Wide Field Camera (WFC). These data were taken over 40 orbits, to which we include an additional 13 orbits on the Phoenix cluster (GO15315) which were previously published \citep{2019ApJ...885...63M}, as well as 2 orbits each from the CLASH survey \citep{2012ApJS..199...25P} for MACS1931 (GO12456) and RXJ1532 (GO12454). Our observing strategy was designed to tune the ramp filters to be centered on the \OII\doublet{3726,3729} doublet at the redshift of each cluster. This choice also avoids contamination from any other strong cooling lines to facilitate interpretation as gas that will eventually form stars. The broadband filters were chosen in such a way as to be free of contamination from the strongest expected emission lines (\OII, \OIII\doublet{4959,5007}, and $H\alpha$) to better model and subtract the underlying continuum from the narrowband observations (described in more detail below in \autoref{subsubsec:cont_subtraction}). A summary of the new and archival data used in this study can be found in \autoref{tab:obs_HST}, where we list the exposure times, proposal IDs, and filters used for each source, including the wavelength each ramp filter was tuned to for observing redshifted \OII emission. 

For each system, we used STScI's {\tt DrizzlePac} software package\footnote{\url{https://drizzlepac.readthedocs.io/en/latest/}} to further process the individual calibrated, flat-fielded exposures. For all visits of a single filter, we used {\tt SExtractor} \citep{1996A&AS..117..393B} to create point source catalogs that were passed into {\tt TweakReg (v1.4.7)} in order to register all exposures to the same World Coordinate System at the sub-pixel level. We then used {\tt AstroDrizzle (v3.1.6)} to remove geometric distortions, correct for sky background variations, flag cosmic rays, and combine individual exposures. This procedure was repeated for both of the broadband filters used for each target. The ramp filter exposures of each system were combined in a similar manner, except with an additional initial step of running {\tt AstroDrizzle} first with only cosmic ray flagging in order to be able to run {\tt SExtractor}, as the raw exposures are usually dominated by cosmic rays due to the filters' low throughputs and narrow fields of view (see ACS instrument science reports\footnote{\url{https://www.stsci.edu/hst/instrumentation/acs/documentation/instrument-science-reports-isrs}} \citealt{2015acs..rept....4L,2015acs..rept....5L}). After registering, cleaning and combining the exposures for each broadband and ramp filter separately, we create point source catalogs and run {\tt TweakReg} again on each of the now combined filter data to align the images across all filters.

\subsubsection{Continuum Subtraction}
\label{subsubsec:cont_subtraction}

In order to do the continuum subtraction, we first compose a spectral energy distribution (SED) for each pixel. This is done by convolving the broadband filter bandpasses $\mathcal{T}_{\rm X}$ (obtained with {\tt pysynphot}) used for each cluster with both a redshifted young (10 Myr) and old (6 Gyr) stellar population spectral template obtained with {\tt STARBURST99} \citep{1999ApJS..123....3L}. The predicted flux from these convolutions is then scaled to match the observed flux in each broadband observation. Because the number of convolved stellar templates is equal to the number of broadband filters, a unique linear combination exists that allows us to convert between blue and red bandpass fluxes ($F_{\rm B}$ and $F_{\rm R}$) to `young' ($\mathcal{Y}$) and `old' ($\mathcal{O}$) template fluxes:

\begin{equation}
    \begin{bmatrix} F_{\rm B} \\ F_{\rm R} \end{bmatrix}
    = 
    \begin{bmatrix}     
        \frac{\int \mathcal{Y} * \mathcal{T}_{\rm B} d\lambda}{\int \mathcal{T}_{\rm B} d\lambda} & 
        \frac{\int \mathcal{O} * \mathcal{T}_{\rm B} d\lambda}{\int \mathcal{T}_{\rm B} d\lambda} \\ 
        \frac{\int \mathcal{Y} * \mathcal{T}_{\rm R} d\lambda}{\int \mathcal{T}_{\rm R} d\lambda} & 
        \frac{\int \mathcal{O} * \mathcal{T}_{\rm R} d\lambda}{\int \mathcal{T}_{\rm R} d\lambda} \\
    \end{bmatrix}
    \begin{bmatrix} c_{\rm \mathcal{Y}} \\ c_{\rm \mathcal{O}} \end{bmatrix}
\end{equation}

\begin{figure*}
\includegraphics[width=\textwidth]{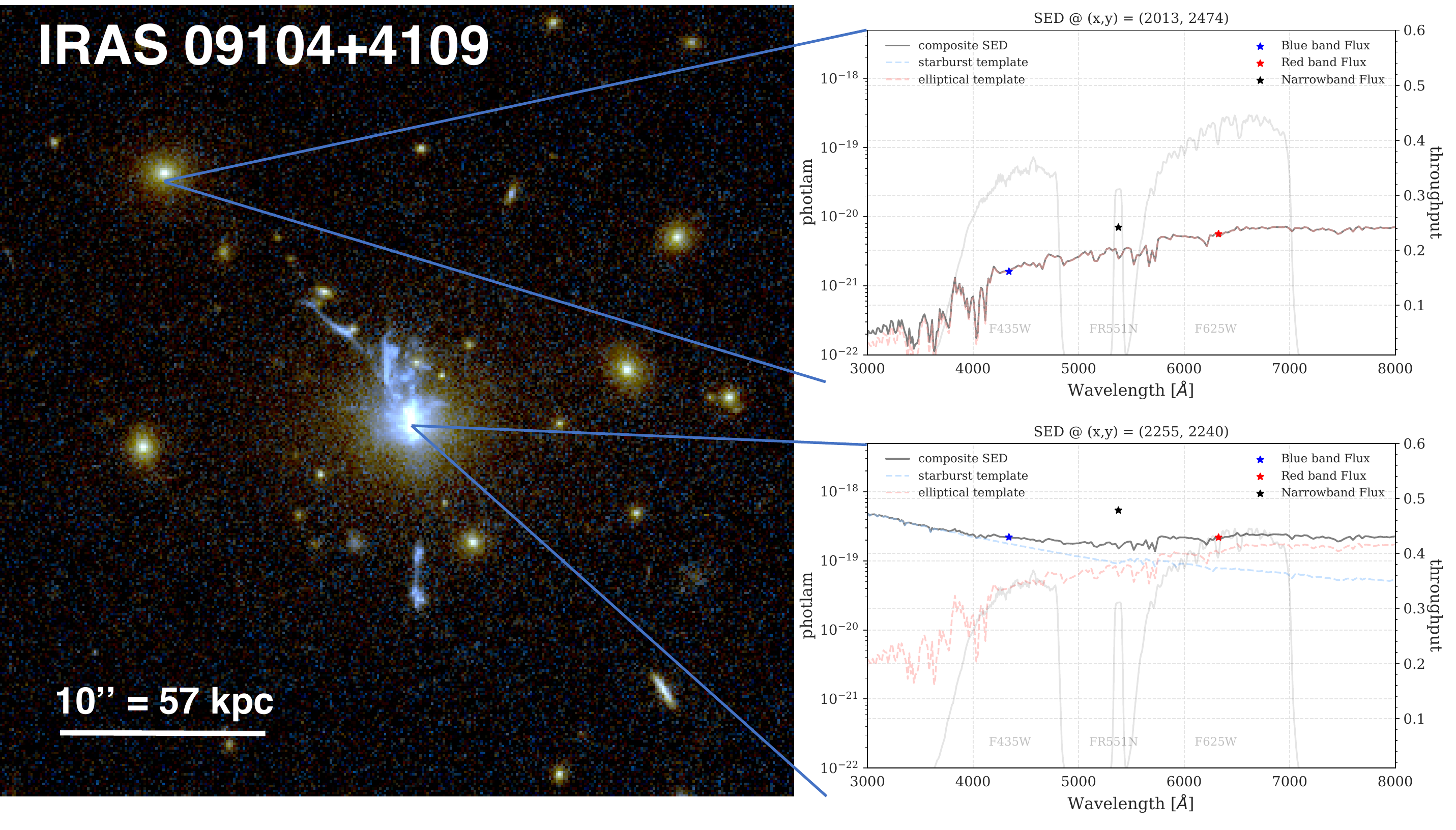}
\caption{Illustration of the pixel-by-pixel spectral energy distribution (SED) fitting procedure we use to produce emission-line only maps (see \autoref{fig:OII_maps}). For each object, we use narrowband observations tuned to the redshifted wavelength of \OII\doublet{3726,3729}, bracketed by images of broadband filters chosen to be free from strong emission lines usually associated with star formation or AGN emission (e.g. \Ha and ${\rm \OIII}$). The flux in a single pixel from the bracketing broadband images is then used to construct an SED and is fit by a linear combination of a 10 Myr starburst and 6 Gyr passive elliptical galaxy template, allowing their normalizations to vary (see top right and bottom right panels, highlighting two distinct pixel fits). The combined, predicted template flux at the wavelength of the narrowband filter is then subtracted from the narrowband image at that pixel, resulting in a continuum-subtracted image after the procedure is run for every common pixel in the aligned images.
\label{fig:SEDfitting}}
\end{figure*}

This equation can be inverted to solve for the vector $\begin{bmatrix}c_{\rm \mathcal{Y}} & c_{\rm \mathcal{O}}\end{bmatrix}^{T}$, the coefficients by which to scale the template spectra to match the observed blue and red broadband fluxes. Because the templates are defined over a broad range of wavelengths, we can integrate the scaled composite young plus old templates over the ramp filter bandpass to estimate the expected narrowband continuum flux ($F_{\rm N}$):

\begin{equation}
\begin{split}
    F_{\rm N} &=
    \begin{bmatrix} 
        \frac{\int \mathcal{Y} * \mathcal{T}_{\rm N} d\lambda}{\int \mathcal{T}_{\rm N} d\lambda} &
        \frac{\int \mathcal{O} * \mathcal{T}_{\rm N} d\lambda}{\int \mathcal{T}_{\rm N} d\lambda}
    \end{bmatrix}
    \begin{bmatrix} c_{\rm \mathcal{Y}} \\ c_{\rm \mathcal{O}} \end{bmatrix}\\
    &=
    \frac{\int \left(c_{\rm \mathcal{Y}}\cdot\mathcal{Y} + c_{\rm \mathcal{O}}\cdot\mathcal{O} \right) * \mathcal{T}_{\rm N} d\lambda}{\int \mathcal{T}_{\rm N} d\lambda}
\end{split}
\end{equation}
Finally, we can subtract this predicted narrowband flux from the observed narrowband flux to get a continuum-subtracted \OII-only flux. This SED fitting and continuum subtraction is done pixel-by-pixel to get \OII-only maps for each of the clusters in our sample. The procedure is illustrated in \autoref{fig:SEDfitting}.

\subsection{X-ray/Chandra}



\begin{table}[t]
  \centering
  \caption{Summary of Archival \textit{Chandra} Observations}
    \begin{tabular}{llc}
    \toprule
    \toprule
    Name & ObsIDs & \multicolumn{1}{l}{t (ks)} \\
    \midrule
    Phoenix  & 13401, 16135, 16545, 19581,  & 548 \\
          & 19582, 19583, 20630, 20631,  &  \\
          & 20634, 20635, 20636, 20797 &  \\
    H1821+643 & 9398, 9845, 9846, 9848 & 86 \\
    IRAS 09104+4109 & 509, 10445 & 84 \\
    Abell 1835 & 6880, 6881, 7370 & 193 \\
    RX J1532.9+3021 & 1649, 1665, 14009 & 108 \\
    MACS 1931.8-2634 & 3282, 9382 & 113 \\
    RBS 797 & 2202, 7902     & 50 \\
    \bottomrule
    \end{tabular}%
  \label{tab:obs_chandra}%
\end{table}%

Archival \textit{Chandra} X-ray data were used for each of the clusters in our sample in order to qualitatively compare to the features seen in the HST \OII maps. A summary of the observations used for this analysis is presented in \autoref{tab:obs_chandra}. All \textit{Chandra} observations were reduced using the \textit{Chandra} Interactive Analysis of Observations (\texttt{CIAO}) \texttt{v4.10} software package, along with \texttt{CALDB v4.8.0}. The observations were reprocessed using \texttt{chandra\_repro}, applying the latest gain and charge-transfer inefficiency corrections, with improved background filtering applied to those observations taken in the \texttt{VFAINT} telemetry mode. Observations from multiple ObsIDs were reprojected, exposure-corrected, and combined using the \texttt{merge\_obs} routine, over an energy range of 0.5--7 keV. Point sources were removed via a wavelet decomposition of the merged images using the \texttt{wavdetect} script, after which periods of high background were excluded using \texttt{lc\_clean}. Blank-sky background files used for background subtraction were obtained using the \texttt{CIAO blanksky} script. These blank-sky files were renormalized to have the same high energy particle rate as the observations over the 9.5--12 keV range, where \textit{Chandra's} effective area is low enough so that any flux is mostly due to the particle background, using the \texttt{blanksky\_sample} script.

\subsubsection{Thermodynamic Profiles}
\label{subsubsec:thermo_rprofiles}

For our spectral analysis, spectra were extracted from concentric annuli centered on the X-ray peak in each system, using broad and fine bin widths, as in, e.g., \citet{2019ApJ...885...63M}. The broader annuli were chosen to contain $\gtrsim$2000 counts to permit precise temperature measurements. The spectra were extracted from the observation and blank-sky background files over the energy range of 0.5--7 keV using \texttt{specextract}, with an identical off-source region also being extracted for both. The spectra from multiple ObsIDs in each bin were fit with the \texttt{XSPEC v12.10.1} spectral fitting package \citep{1996ASPC..101...17A} using the Cash statistic \citep{1979ApJ...228..939C} and Levenberg-Marquardt algorithm. Each on-source spectrum was fit using a combined spectral model \texttt{PHABS*(APEC + BREMSS) + APEC}, where the first \texttt{APEC} component is to model thermal emission from the optically thin plasma in the ICM \citep{2001ApJ...556L..91S} along with Galactic photoelectric absorption using \texttt{PHABS}. We further model the background with a second \texttt{APEC} component (fixed at $kT = 0.18$ keV, $Z = Z_{\odot}, z=0$) to model soft Galactic X-ray emission, as well as a \texttt{BREMSS} model (fixed at $kT = 40$ keV) to model unresolved point sources \citep[e.g.][]{2019ApJ...885...63M}. These two background components are jointly fit with the off-source spectra, which are only modeled with \texttt{PHABS*BREMSS + APEC}, with normalizations scaled by the extraction area of the on-source spectra. Galactic H\textsc{\lowercase{I}} column densities ($N_{\rm H}$) for each system were taken from \citet{2005A&A...440..775K}, redshifts were fixed to their literature values, and metallicities for the on-source \texttt{APEC} components were fixed at $Z = 0.3 Z_{\odot}$. Only \texttt{APEC} temperatures and normalizations were allowed to vary.

The resulting coarse temperature profiles extracted along each annular bin were fit with the analytical model of \citet{2006ApJ...640..691V}:

\begin{equation}
    T_{\rm 3D}(r) = T_0  \frac{\left(r/r_{\rm cool}\right)^{\alpha} + T_{\rm min}/T_0}{\left(r/r_{\rm cool}\right)^{\alpha} + 1}  \frac{\left(r/r_{\rm t}\right)^a}{[1 + \left(r/r_{\rm t}\right) ^b]^{c/b}}
\end{equation}

where $a$, $b$, and $c$ model the outer regions of the profile with a flexible broken power-law. The profile transitions to an inner cool core component at around $r_t$, with the cool core defined by $T_0$, $T_{\rm min}$, $r_{\rm cool}$, and $\alpha$.
All fitting parameters are initialized to the average values found in \citet{2006ApJ...640..691V}. This 3D temperature was projected along the line of sight
and over the width of each annular bin, and then fit to the observed profile using the \texttt{python} package \texttt{LMFIT} \citep{2014zndo.....11813N}. The corresponding best fit temperature profile was then interpolated along a finer radial binning from which we extracted another profile. In these finer spectral fits, we keep the temperatures fixed at these interpolated values. Only the \texttt{APEC} normalization is allowed to vary in order to reduce the degrees of freedom and more precisely constrain the density profile. The \texttt{APEC} normalization $\eta$ was converted to an emission measure (EM) profile using EM(r) $\equiv \int n_{\rm e} n_{\rm p} dV  = \eta \times 4\pi \times 10^{14} \left[ D_{\rm A} (1+z) \right]^2$, where $n_{\rm e}$ and $n_{\rm p}$ are the electron and proton number densities, and $D_{\rm A}$ is the angular diameter distance at the cluster redshift. The EM profile was fit with the following model from \citet{2006ApJ...640..691V}:

\begin{multline}
    n_{\rm e} n_{\rm p} (r) = n_0^2  \frac{\left(r/r_{\rm c}\right)^{-\alpha}}{\left[1+\left(r/r_{\rm c}\right)^2\right]^{3\beta - \alpha/2}}  \frac{1}{\left[1 + \left(r/r_{\rm s}\right)^{\gamma}\right]^{\epsilon/\gamma}} \\
    + \frac{n_{02}^2}{\left[1+\left(r/r_{\rm c2}\right)^2\right]^{3\beta_2}}
\end{multline}

which is a modified double-beta model with a cusp, rather than a flat core (defined by $n_0$, $r_{\rm c}$, $\alpha$, $\beta$), a steeper outer profile slope (defined by $r_{\rm s}$, $\gamma$, and $\epsilon$), and a cool core component (defined by $n_{\rm 02}$, $r_{\rm c2}$, and $\beta_2$). In our fits, $\gamma =3$ remains fixed, and all other parameters are allowed to vary and initialized to typical parameters found in \citet{2006ApJ...640..691V}. This 3D model was also projected along the line of sight.
From this fit, the electron number density was calculated assuming abundances from \citet{1989GeCoA..53..197A} for a fully ionized 0.3$Z_{\odot}$ abundance plasma, such that $n_{\rm e}/n_{\rm p} = 1.2$.

Additional thermodynamic profiles are derived analytically from the fitted density and temperature profiles.
We calculate profiles for the total pressure ($P = (n_{\rm e} + n_{\rm p}) k_{\rm B}T$), pseudo-entropy ($K = k_{\rm B} T n_{\rm e}^{-2/3}$), cooling time ($t_{\rm cool} = \frac{3}{2}\frac{(n_{\rm e} + n_{\rm p}) k_{\rm B} T}{n_{\rm e} n_{\rm p} \Lambda(k_{\rm B}T, Z)}$), and freefall time ($t_{\rm ff} = \sqrt{2r/g}$). We used the cooling function $\Lambda(k_{\rm B}T, Z)$ from  \citet{1993ApJS...88..253S}, as parameterized by \citet{2001ApJ...546...63T} (see also \citealt{2009ApJ...703...96P,2008MNRAS.384..251G}). The gravitational acceleration used in calculating the freefall time was obtained by modeling the cluster potential with the sum of an isothermal sphere at small radii and a Navarro-Frenk-White (NFW) profile \citep{1997ApJ...490..493N} at larger radii, assuming a velocity dispersion of $\sigma_v = 350$ km s$^{-1}$ typical of BCGs \citep[e.g.][]{2007MNRAS.379..867V,2020ApJ...891..129S}.
We compare the profiles of each of our sources to measurements from the literature to confirm agreement between our thermodynamic modeling. Literature sources we referenced for comparison include: \citet{2019ApJ...885...63M} for Phoenix, \citet{2010MNRAS.402.1561R} for H1821, \citet{2012MNRAS.424.2971O} for IRAS09104, \citet{2009ApJS..182...12C} for Abell 1835, RXJ1532 and MACS1931, as well as \citet{2011MNRAS.411.1641E} for MACS1931, and \citet{2011ApJ...732...71C} and \citet{2012ApJ...753...47D} 
for RBS797. Data for the Phoenix cluster and H1821+643 were obtained from M. McDonald and H. Russell, respectively (via private communication), and profiles were not extracted independently here, only modeled analytically, due to the complicated nature of the AGN's contribution in these systems.

\begin{figure*}[]
\centering
\includegraphics[width=\textwidth,height=2.5in]{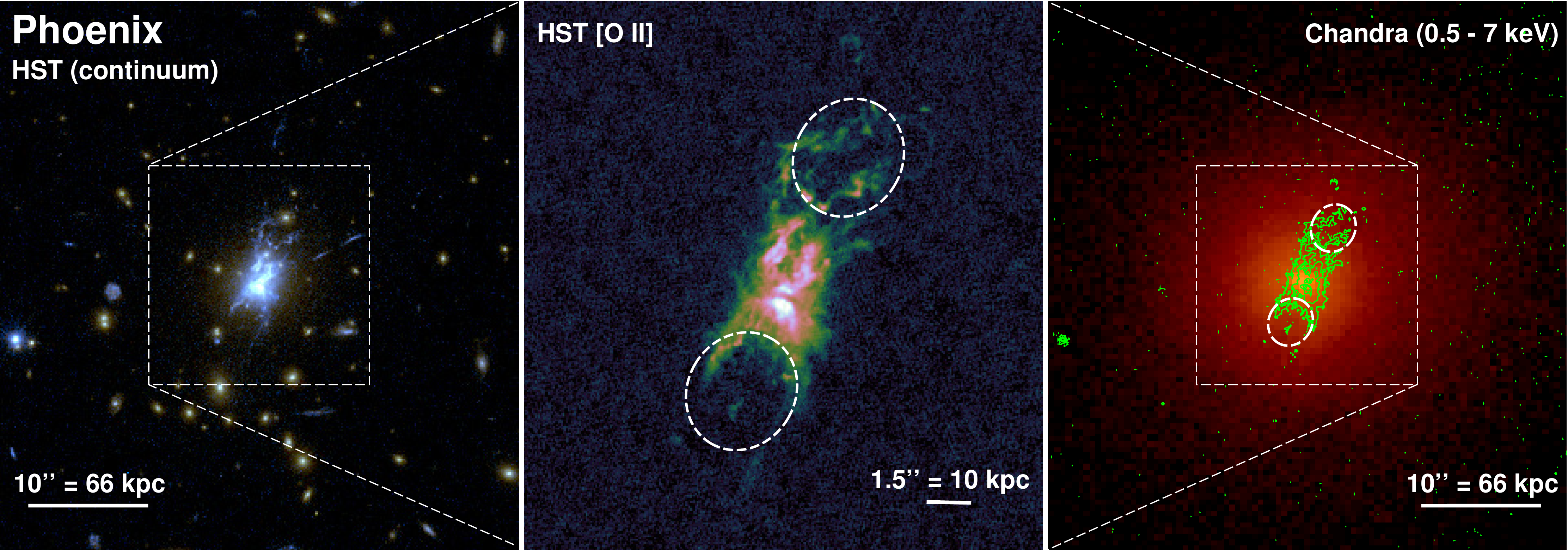}
\includegraphics[width=\textwidth,height=2.5in]{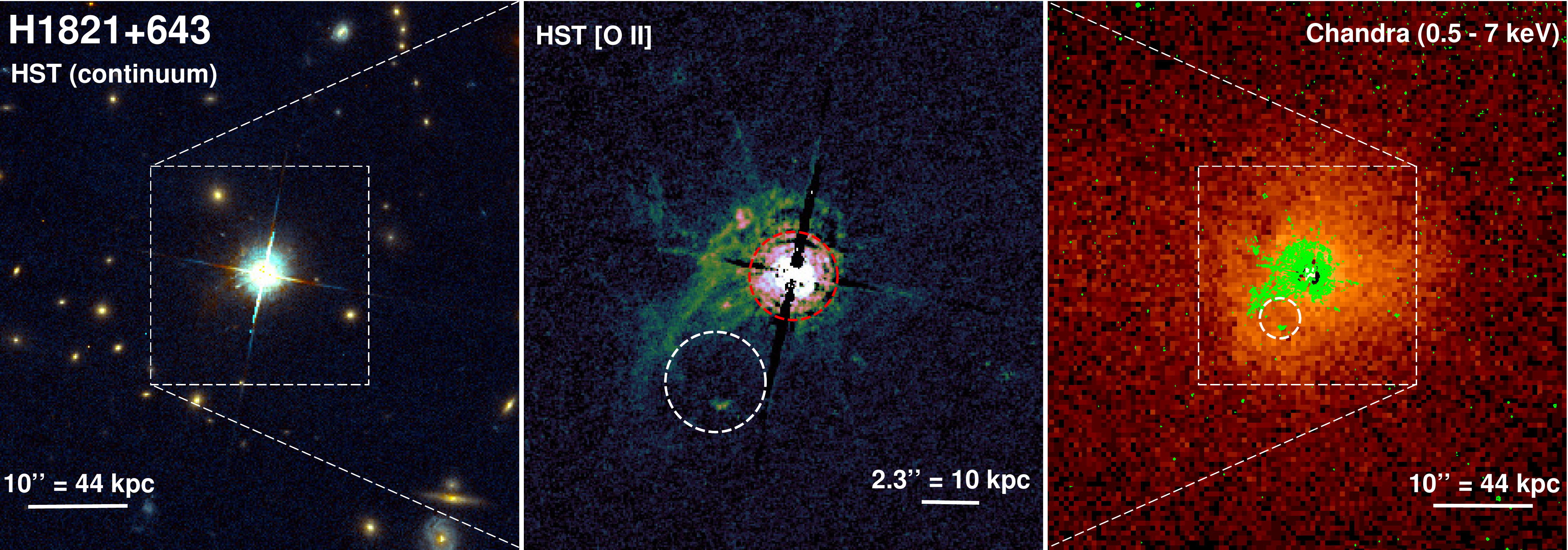}
\includegraphics[width=\textwidth,height=2.5in]{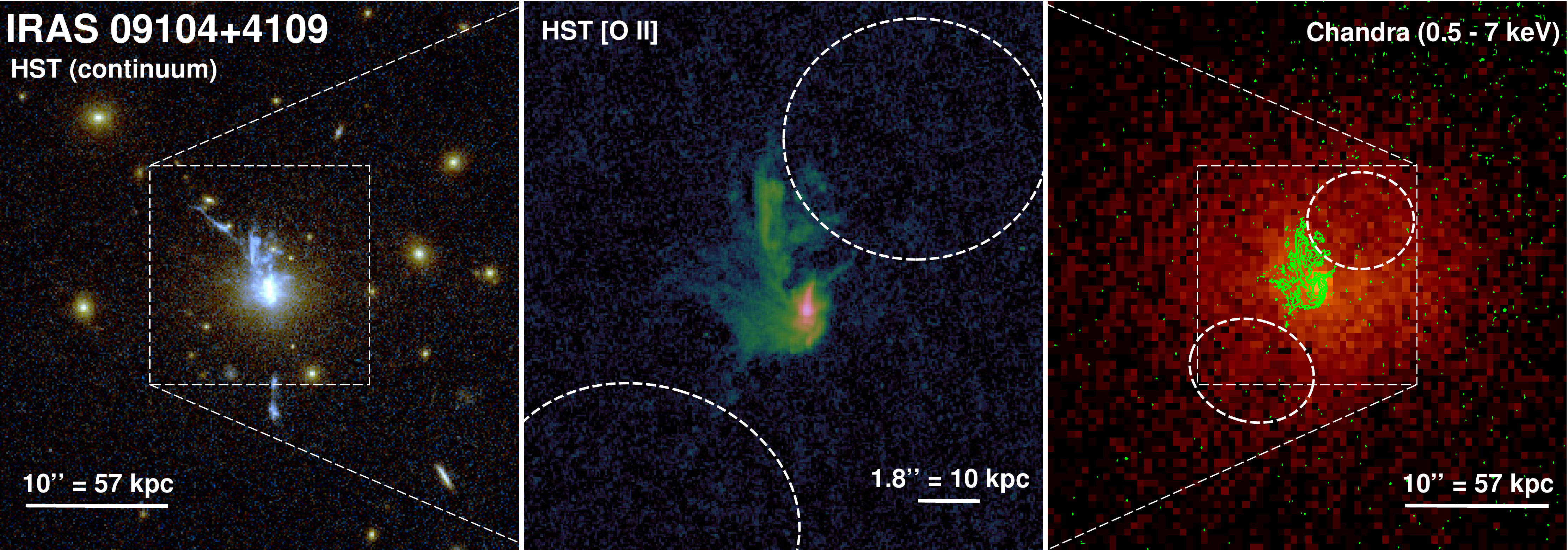}
\caption{Continuum-subtracted \OII maps of the BCGs in our sample, including Phoenix, H1821+643, IRAS 09104+4109, Abell 1835, RX J1532.9+3021, MACS 1931.8-2634, and RBS 797. In the \textit{\textbf{left}} panel is a combination of broadband HST images using the filters listed in \autoref{tab:obs_HST}, as well as the continuum predicted by the procedure described in \autoref{subsubsec:cont_subtraction} and \autoref{fig:SEDfitting}. The \textit{\textbf{middle}} panel is a zoom in of the central \OII emitting nebulae with the continuum subtracted. Elliptical dashed regions denote the locations of known cavities from the literature that can also be identified from the \textit{Chandra} X-ray images shown in the \textit{\textbf{right}} panel, with overlaid \OII contours in green. The region of H1821 within the red-dashed circle was masked to exclude possible contamination from the bright central quasar to the SFR measurement. \label{fig:OII_maps}}
\end{figure*}

\addtocounter{figure}{-1}
\begin{figure*}[]
\centering
\includegraphics[width=\textwidth,height=2.5in]{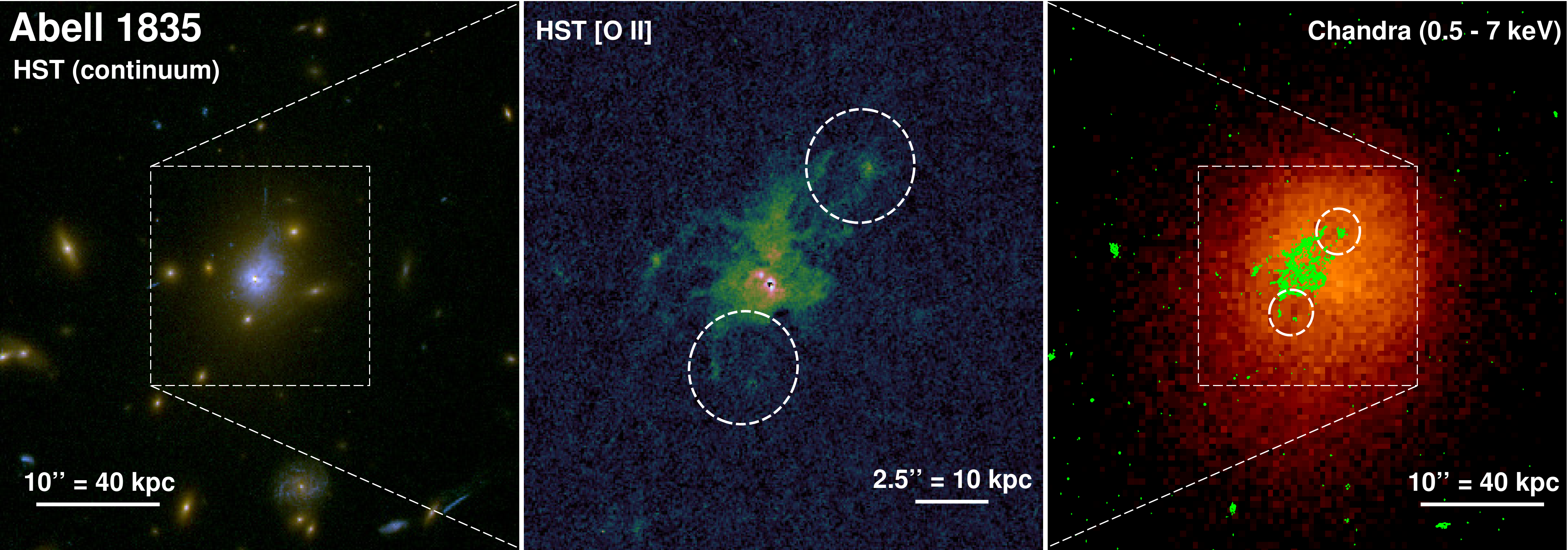}
\includegraphics[width=\textwidth,height=2.5in]{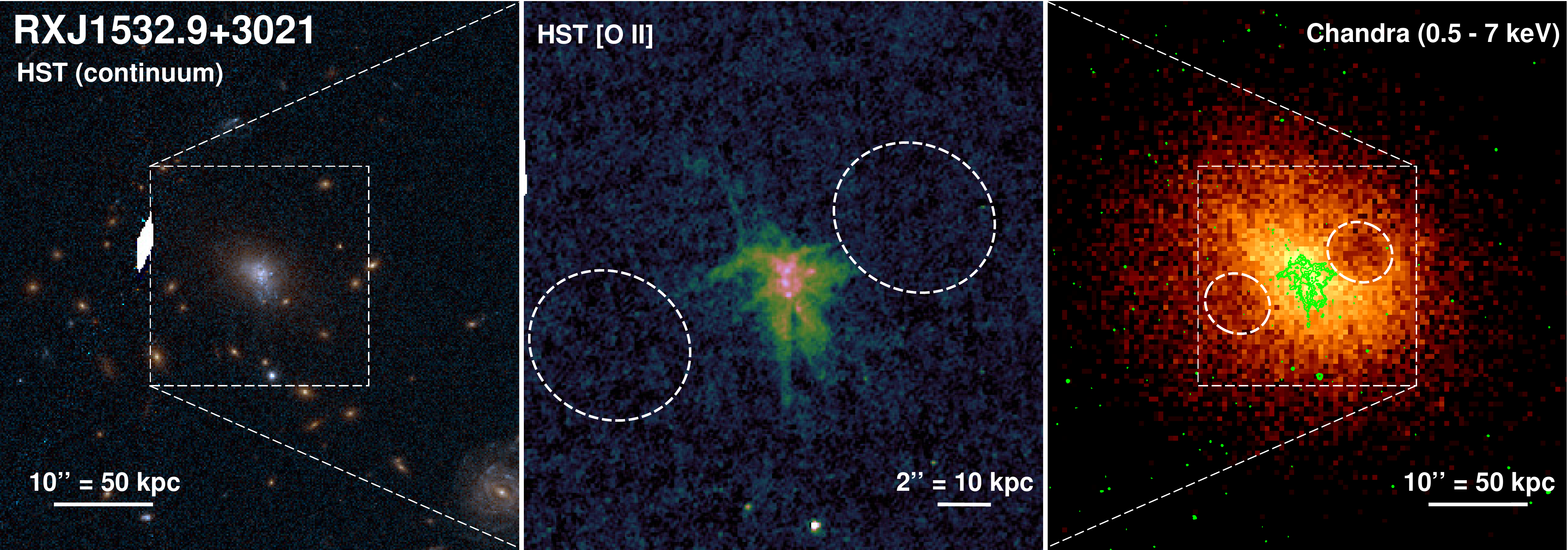}
\includegraphics[width=\textwidth,height=2.5in]{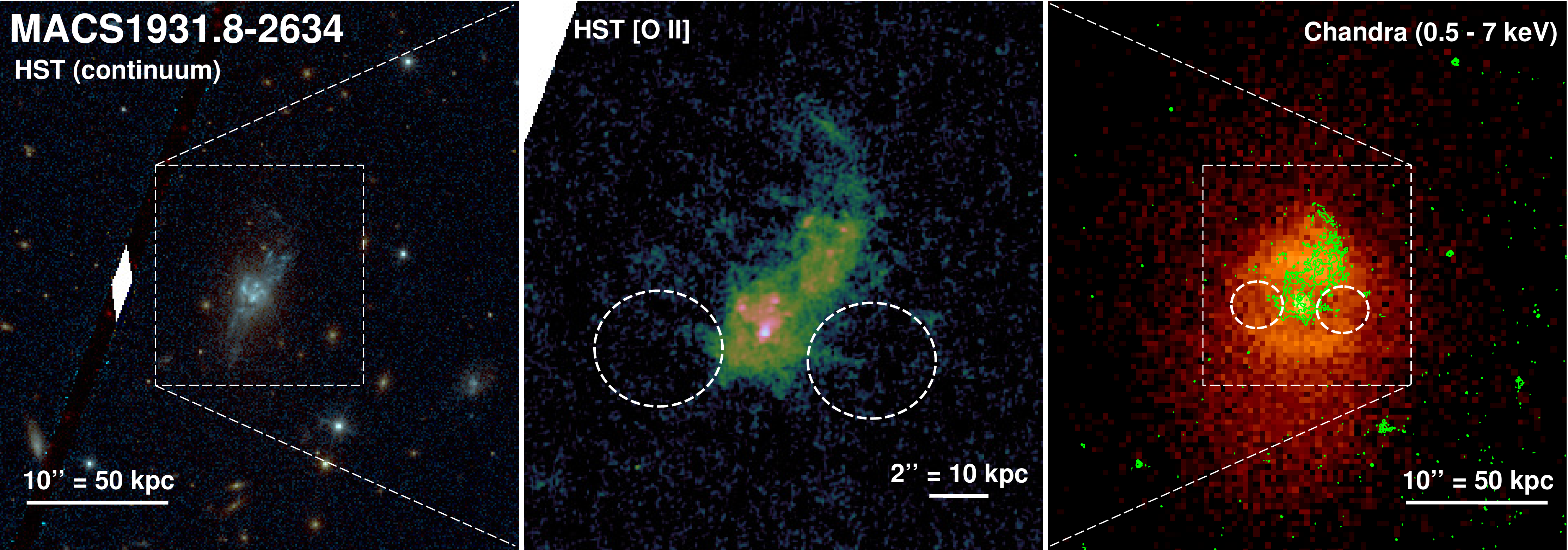}
\caption{(continued)}
\end{figure*}

\addtocounter{figure}{-1}
\begin{figure*}[]
\centering
\includegraphics[width=\textwidth,height=2.5in]{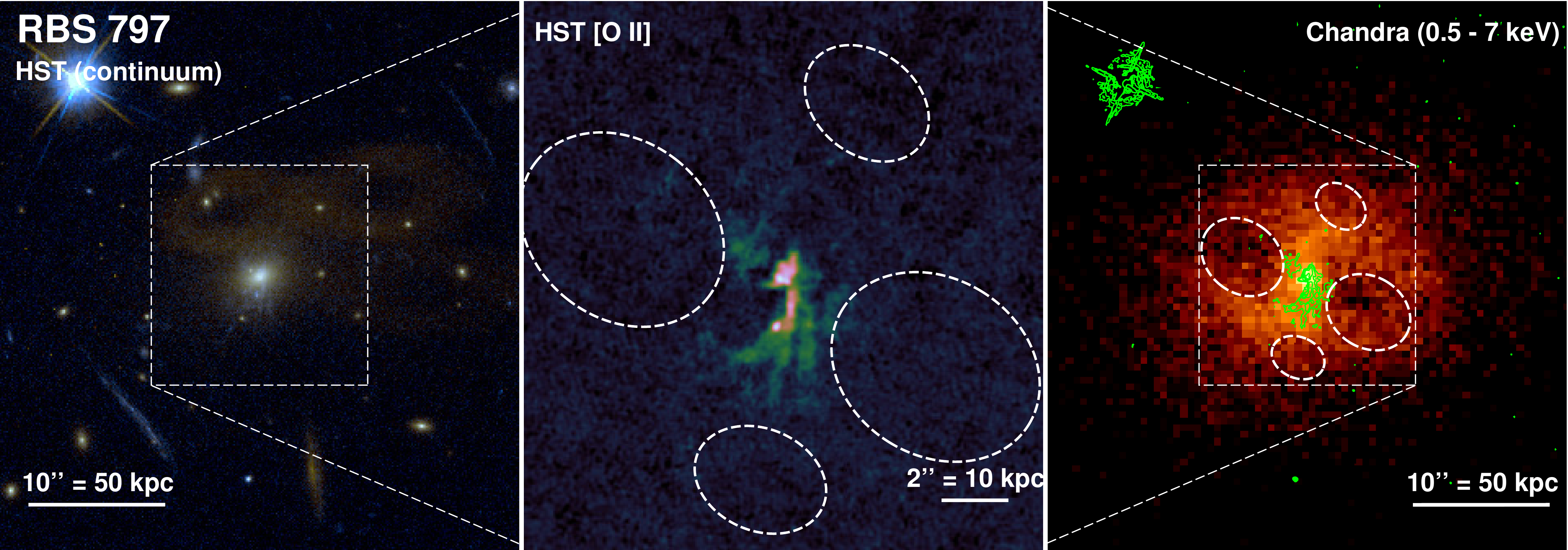}
\caption{(continued)}
\end{figure*}

\section{Results} \label{sec:results}

\subsection{{\rm \OII}Maps} 
\label{subsec:OII_maps}

Our new, reduced broad- and narrow-band HST observations can be found in \autoref{fig:OII_maps}. 
The continuum-subtracted \OII maps in the middle panels reveal with remarkable detail the intricate morphology of the warm ($T {\sim} 10^4$ K) ionized gas within each of the BCGs in our sample that will eventually form stars.
These maps represent the highest angular resolution view of the star-forming material at optical wavelengths in each of these strongly cooling clusters to date (with the exception of Phoenix which was already presented in \citealt{2019ApJ...885...63M}).
Each of these nebulae are extended on the order of tens of kpc in projection, ranging from $R_{\rm \OII} =$ 23 -- 60 kpc, making them among the most extended emission line nebulae \citep[see e.g.][]{2010ApJ...721.1262M,2016MNRAS.460.1758H}. The most extended of these filaments is found in Phoenix, where the \OII filaments reach up to 60 kpc \citep{2019ApJ...885...63M}, similar to the well-studied \Ha filament system found in the extremely deep observations of NGC1275 at the center of the nearby ($z=0.018$) Perseus cluster \citep{2001AJ....122.2281C,2003MNRAS.344L..48F}, with $R_{\rm H\alpha} \approx$ 40 kpc.

The morphologies of the \OII maps appear to be extending either in various directions (as is the case in Phoenix, Abell 1835, and RXJ1532) or in predominantly one direction (as in H1821, IRAS09104, MACS1931, and RBS797). The cause of these different morphologies may be investigated by looking to observations from other wavelengths. For instance, in the Phoenix cluster, the X-shaped  morphology of the pairs of filaments to the North and South are coincident with the rims of X-ray cavities carved out by radio bubbles, as detailed in \citet{2019ApJ...885...63M} (see \autoref{fig:OII_maps}). Cold molecular gas coincident with \Ha filaments, as measured by ALMA \citep{2017ApJ...836..130R}, suggest that in a few systems buoyantly-rising radio bubbles are promoting cooling in their wake as they climb out of their deep cluster potentials in a few systems \citep[e.g.][]{2008A&A...477L..33R,2010MNRAS.406.2023P,2016ApJ...830...79M}. This picture has been similarly suggested as the mechanism for cooling in Abell 1835, where the \OII filaments presented here for the first time are also coincident with molecular gas rising behind the location of X-ray cavities \citep{2014ApJ...785...44M}. For each of the clusters in our sample, we can see the morphology of the \OII overlaid on top of the X-ray maps of the cluster cores, along with the location of known X-ray cavities from the literature in \autoref{fig:OII_maps}. Cool gas in some clusters appears to be extended in mostly one direction, in some cases completely perpendicular to the direction of the known X-ray cavities, as in MACS1931, where there is also molecular gas that traces the \OII and {\Ha}+ ${\rm \NII}$ morphology \citep{2019ApJ...879..103F}. A possible explanation for the triggering of star formation that is not stimulated by rising radio bubbles could be a recent gas-rich merger that deposited the cooling material or created turbulence to promote existing material to cool as per the CCA model.
Clearly, the location of bubbles does not necessarily predict the direction of cooling. Furthermore, even in the case of Phoenix, where there is obvious agreement between the position angles of cavities and star-forming filaments, the \OII filaments extend even beyond the maximum radius of the cavities, implying that radio bubbles do not tell the whole story. We will discuss the implications of the presence of bubbles on the development of thermal instabilities further in \autoref{subsubsec:R_OII_vs_bubbles}.

\vspace{10mm}

\subsection{Star Formation Rates}
\label{subsec:SFRs}
One of the main goals of our new HST observations was to secure more precise star formation rates (SFRs) by being able to more faithfully isolate the morphology of star-forming filaments and exclude likely regions of high AGN contamination. 
To that end, we extract an \OII flux $f_{\rm \OII}$ from 
polygonal apertures designed to encompass all of the flux while maximizing the signal-to-noise for each of the \OII nebulae.
We applied an intrinsic extinction correction to each of our rest-frame \OII flux measurements based on the $E(B-V)$ measurements by \citet{1999MNRAS.306..857C}, assuming $R_{\rm V} = 3.1$. In the case of Phoenix, we used an $E(B-V) = 0.24 \pm 0.10$ from \citet{2014ApJ...784...18M}. \citet{1999MNRAS.306..857C} measured specific reddening values for A1835 of $E(B-V) = 0.38\pm0.04$ and for RXJ1532 of $E(B-V) = 0.21\pm0.03$. For the rest of our sample, we take the distribution of reddening values from the entire sample of \citet[][Table 4, col. 7]{1999MNRAS.306..857C}, and use the 1st, 2nd, and 3rd quartiles to apply a reddening of $E(B-V) = 0.27^{+0.16}_{-0.07}$. Following this intrinsic extinction correction, we also apply a Galactic extinction correction at the redshifted \OII wavelengths using the values listed in the NASA/IPAC Extragalactic Database (NED), which are based on Sloan $r'$ band data from \citet{2011ApJ...737..103S}. Following these corrections, we converted our corrected \OII fluxes to SFRs using the scaling relations found in \citet{2004AJ....127.2002K}:

\begin{equation}
    SFR_{\rm \OII} = 6.58 \times 10^{-42} L_{\rm \OII} ~~{\rm M}_{\odot}~ {\rm yr}^{-1}
\end{equation}
where $L_{\rm \OII} = f_{\rm \OII} \cdot 4 \pi D_{\rm L}(z)^2$ is units of erg s$^{-1}$, and $D_L(z)$ is the redshift-dependent luminosity distance. These new \OII SFRs can be found in \autoref{tab:sfr_new}. We find good agreement between our new measurements and the literature, particularly in the case of Phoenix \citep{2019ApJ...885...63M}, as well as MACS1931 and RXJ1532 \citep{2017ApJ...846..103F}, IRAS09104 \citep{2013A&A...549A.125R}, and Abell 1835. Previous SFR values for RBS797 based on UV fluxes \citep{2011ApJ...732...71C} are much lower than our \OII SFR measurements, likely due to no extinction correction being made.



Special care had to be taken to extract an accurate $f_{\rm \OII}$ for H1821, as the bright quasar produced strong negative residuals in the continuum-subtraction due to the diffraction spikes, as seen in \autoref{fig:OII_maps}. These negative residuals were masked in our extraction region. Also, we might expect that some of the ionized \OII emission is due to radiation from the quasar and not from star formation. To account for this, we initially tried removing a central region up to a radius defined by the Str{\"o}mgren sphere: 

\begin{equation}
    r_{Str\ddot{o}mgren} = \left[ \frac{3 N_{\rm i}}{4 \pi \alpha_{\rm B} \langle n_{\rm e}\rangle^2} \right]^{1/3}
\end{equation}
where $\alpha_{\rm B} = 2.6 \times 10^{-13}$ cm$^3$ s$^{-1}$ is the coefficient for case B recombination (i.e. only \textit{net} recombinations, thus excluding transitions directly to the ground state), $\langle n_{\rm e}\rangle=300$ cm$^{-3}$ is the average electron density of cool line-emitting gas in cluster cores \citep[e.g.][]{2012ApJ...746..153M}, and $N_{\rm i}$ is the ionizing photon rate. The ionizing photon rate was estimated by finding a best-fit relation of $\log N_{\rm i} = 1.05 \log L_{\rm bol} + 7.46$ between the bolometric luminosity $L_{\rm bol}$ and ionizing photon rate for the quasar sample in \citet{1994ApJS...95....1E} (see their Table 14).
Using a bolometric luminosity of $L_{\rm bol} = 2\times10^{47}$ erg s$^{-1}$ for H1821 \citep{2010MNRAS.402.1561R}, we find an approximate ionizing photon rate of $N_{\rm i} \approx 10^{57}$ s$^{-1}$ and a corresponding Str{\"o}mgren sphere radius of about 0.7 kpc = 0\farcs16. The amount of quasar contamination to the \OII flux within such a small radius is negligible, and even more so for the rest of the systems in our sample. However, given that this Str{\"o}mgren sphere calculation presumes a homogeneous medium, and that the filling factor of the line-emitting gas is likely low, the \textit{effective} Str{\"o}mgren sphere radius should be larger. In addition, the Airy diffraction pattern induced by the bright quasar in H1821 also leaves behind positive residuals which cannot be included in the SFR calculation. Instead, in lieu of the complexity of modeling the HST PSF, we calculate a radius of 1\farcs8 to exclude from the center of this source (red dashed circle in \autoref{fig:OII_maps}), which corresponds to an encircled energy fraction of $\sim 99.8\%$ of an on-axis point source. As a result, the SFR we measure for H1821 can be considered a lower limit, though the extent of the missing flux from young stars is uncertain as much of the ionization from this masked region may come from the quasar.


\begin{figure*}[t]
\includegraphics[width=\textwidth]{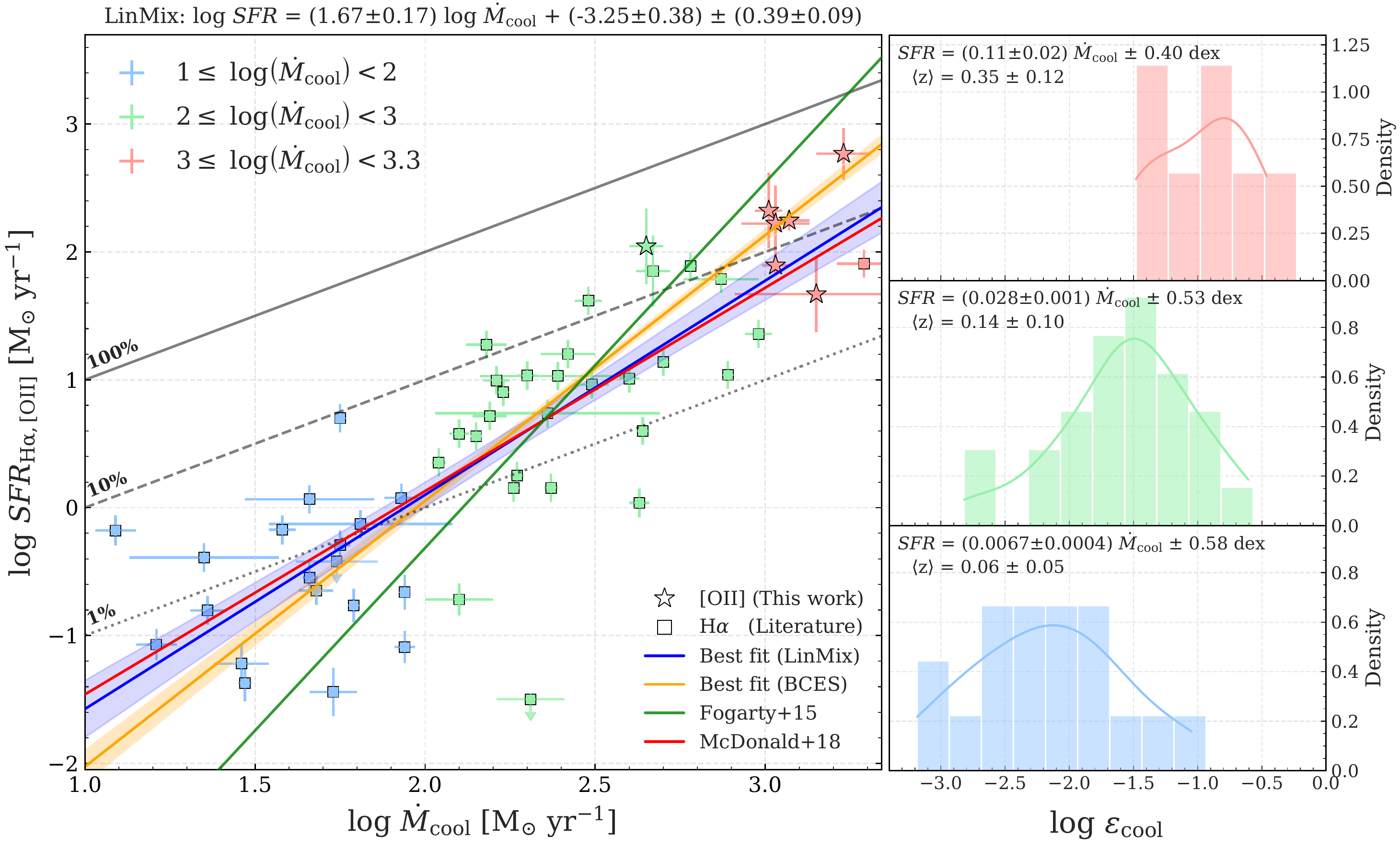}
\caption{\textit{\textbf{Left}}: Maximal ICM cooling rates ($\mdot = M_{\rm gas}(r < r_{\rm cool})/t_{\rm cool}$) vs. star formation rates (SFR) based on new \OII observations for our sample (starred points; see \autoref{tab:sfr_new}), along with homogenized \Ha values from the literature (squares; see \autoref{tab:Halpha_corrected}). Curves of constant cooling efficiency ($\epsilon_{\rm cool} \equiv SFR/\dot{M}_{\rm cool}$) are plotted as diagonal black lines, from 1\% to 100\%. Our best fit relation between SFR and ${\mdot}$ obtained from a robust Bayesian analysis is quoted at the top and shown as a solid blue line with shaded $1\sigma$ uncertainty band, along with the best fit relations from \citet[][solid green line]{2015ApJ...813..117F} and \citet[][solid red line]{2018ApJ...858...45M}, and a BCES orthogonal fit (orange line). The steeper-than-unity slope in all these relations suggests that cooling efficiency increases with ${\mdot}$, implying a \textit{gradual saturation of AGN feedback}, likely tied to an increasing fraction of feedback energy output being radiative as opposed to mechanical. Interestingly, the three starred data points with the highest cooling efficiency are also quasar-hosting systems (Phoenix, H1821, IRAS09104). Points are color coded to correspond to different cooling rate bins, as described in the legend in the top left of this panel. \textit{\textbf{Right}}: Distributions of cooling efficiency after binning the points from the \textit{left} panel by cooling rate and using the same color coding. 
Solid curves overlaid on each histogram are smooth probability density functions calculated non-parametrically via kernel density estimation. It is visually straightforward to see that with higher cooling rate bins, the mean cooling efficiency increases significantly. A slight redshift trend may be due to selection bias, but further analysis is left for a future study.
\label{fig:sfr_vs_mdot}}
\end{figure*}

In addition to our sample of 7 clusters, we compare their SFRs to those of other cool core clusters from the literature. 
In particular, we searched for systems with available \Ha flux measurements of strongly cooling groups and clusters of galaxies, as \Ha 
is a more readily available measurement for lower redshift clusters in the current literature. We compiled the \Ha fluxes/luminosities from \citet{2016MNRAS.460.1758H,2010ApJ...721.1262M,2016AA...588A..68G,2009AA...495.1033B,2018MNRAS.481.4472L,2009ApJS..182...12C,1992ApJ...385...49D,1999MNRAS.306..857C,2010MNRAS.401..433W,1989ApJ...338...48H}, prioritizing first the sources with tunable filter or integral field spectroscopy, and then those with long-slit spectroscopy. For sources that quoted a {\Ha}+ ${\rm \NII}$ flux, we converted to a pure \Ha flux by using the ratio of {\Ha}/ ({\Ha}+ ${\rm \NII}$) = 0.76 \citep[e.g.][]{2010ApJ...721.1262M}. We homogenized each of these literature flux measurements to conform to our same chosen cosmological parameters. In almost every case, a Galactic extinction correction had already been applied, after which we applied an intrinsic correction according to the distribution of $E(B-V)$ values from \citet{1999MNRAS.306..857C}, which becomes the largest source of uncertainty in each of these measurements. Finally, we attempt to remove the contamination to the \Ha flux from evolved stars (e.g. planetary nebulae, supernova remnants, AGB and HB stars, etc.). This contamination is likely to significantly contribute to the SFR especially for the weakly cooling clusters. To account for this contamination, we follow the same procedure described in \citet{2021ApJ...908...85M}, where more details can be found.

\begin{table}[t]
  \caption{New \OII SFRs For Our Sample}
  \centering
    \begin{tabular}{lcc}
    \toprule
    \toprule
    Name & $\log_{10}{\rm SFR}_{\rm \OII}$ & $\log_{10}\dot{M}_{\rm cool}$ \\
    & (M$_{\odot}$ yr$^{-1}$) & (M$_{\odot}$ yr$^{-1}$) \\
    \midrule
    Phoenix & 2.77$\pm$0.20 & 3.23$\pm$0.08 \\
    H1821+643 & 2.04$\pm$0.30 & 2.65$\pm$0.05 \\
    IRAS 09104+4109 & 2.32$\pm$0.30 & 3.01$\pm$0.04 \\
    Abell 1835 & 2.25$\pm$0.08 & 3.07$\pm$0.06 \\
    RX J1532.9+3021 & 1.89$\pm$0.06 & 3.03$\pm$0.04 \\
    MACS 1931.8-2634 & 2.22$\pm$0.30 & 3.03$\pm$0.10 \\
    RBS 797 & 1.67$\pm$0.30 & 3.15$\pm$0.24 \\
    \bottomrule
    \end{tabular}%
  \label{tab:sfr_new}%
\end{table}%

The SFRs for our extreme cooling sample, as well as the reference samples listed above, are plotted against the X-ray cooling rates ($\mdot$; compiled by \citealt{2018ApJ...858...45M}) in \autoref{fig:sfr_vs_mdot}, and can also be found in \autoref{tab:sfr_new} and \autoref{tab:Halpha_corrected}, including the intermediate homogenization of the literature luminosity values. The classically inferred ICM cooling rates here are defined as $\mdot = \frac{M_{\rm gas}(r < r_{\rm cool})}{t_{\rm cool}}$, where $r_{\rm cool}$ is the radius where the cooling time is ${<} 3$ Gyr, and $t_{\rm cool}$ is 3 Gyr. This value of the cooling radius/timescale is chosen to more closely probe the cluster core where cooling actually occurs. There is typically good agreement between these estimates of cooling rates and luminosity-based rates, while other choices of the cooling radius (e.g. $r_{\rm cool}$ at $t_{\rm cool} = 7.7$ Gyr) are essentially arbitrary but useful conventions for comparing to the literature. Spectroscopic based cooling rates are more accurate measurements of how much gas is actually cooling, which is typically less than that inferred by ``classical'' (i.e. maximal) cooling rates and would tend to bring $\mdot$ measurements within an order of magnitude or less of the SFR measurements. However, spectroscopic cooling rates are much harder to measure and usually result in upper limits for cooling clusters. Our choice of cooling rate method is useful as an upper limit to the total rest mass of gas that could potentially cool, allowing energy budget considerations that can more conveniently reveal the impact of heating from AGN feedback. 

Both SFR and $\mdot$ for \autoref{fig:sfr_vs_mdot} range over several orders of magnitude, and we see that our sample of starbursting BCGs lies at the extreme ends of both SFR and $\mdot$, with much higher cooling efficiencies (defined as the ratio $\epsilon_{\rm cool} = SFR / \dot{M}_{\rm cool}$) than the reference systems. 
For the reference systems, we find a combined average cooling efficiency of $1.3\pm0.1$\%, with a log-normal scatter of 0.65 dex, demonstrating the well-known two orders of magnitude suppression of the cooling flow problem. 
In contrast, our sample of extreme cooling clusters ($N=7$) has an average cooling efficiency of $20\pm13$\%. 
When binning the datapoints by cooling rate, one can readily see that systems in higher cooling rate bins have higher average cooling efficiencies, as shown by the colored histograms corresponding to the same colored datapoints in the scatterplot in \autoref{fig:sfr_vs_mdot}.
Motivated by this trend, we fit the SFR vs $\mdot$ plot with a total least squares regression using the \texttt{python} software package \texttt{LinMix} \citep{2007ApJ...665.1489K}. \texttt{LinMix}\footnote{\url{https://linmix.readthedocs.io/en/latest/index.html}} uses a hierarchical Bayesian approach to linear regression for data with both $x$ and $y$ errors, as well as robustly accounting for censored data (i.e. upper limits). We find a relation between cooling and star formation of ${\rm \log(SFR)} = (1.67\pm0.17) \, {\rm \log(\mdot)} + (-3.25\pm0.38)$ with an even lower intrinsic scatter of $0.39\pm0.09$ dex compared to the 0.65 dex from averaging over the entire sample and assuming a constant cooling efficiency. This relation tells us that at higher cooling rates, 
star formation proceeds with greater efficiency, consistent with \citet{2018ApJ...858...45M} and \citet{2015ApJ...813..117F}, though the latter found a steeper relation.

The steeper slope found by \citet{2015ApJ...813..117F} may be attributed to a few different factors. \autoref{fig:sfr_vs_mdot} includes a large number of cool cores with $\mdot < 100$ {\msunperyear} and a measured SFR, whereas \citet{2015ApJ...813..117F} does not. Additionally, the \citet{2015ApJ...813..117F} relation was based on a slightly different cooling rate definition than ours, where they measure $\mdot$ within a fixed 35 kpc aperture as well as one where  {\tcool/\tff} $<20$.
Furthermore, the CLASH sample considered in \citet{2015ApJ...813..117F} had a mean redshift of $\langle z \rangle = 0.392$ and a complicated selection function, while the linear fit in \autoref{fig:sfr_vs_mdot} was performed over many more systems, spanning from $0 \lesssim z \lesssim 0.5$ (${\sim} 5$ Gyr in lookback time), with an average redshift of $\langle z \rangle = 0.183$. 
For comparison, we perform a BCES orthogonal fit to our data and find a best-fit relation of ${\rm \log(SFR)} = (2.08\pm0.14) \, {\rm \log(\mdot)} + (-4.12\pm0.38)$, in closer agreement with the slope of \citet{2015ApJ...813..117F}. We note that this steeper slope does not affect the interpretation of our results in the discussion below (\autoref{sec:discussion}).
We see in the right panel of \autoref{fig:sfr_vs_mdot} a slight redshift trend in the cooling efficiency histograms, where higher redshift systems on average have higher $\epsilon_{\rm cool}$. While not highly significant, this redshift trend could likely be due to observational bias, where it is harder to find less massive systems at higher redshifts. Alternatively, if real this trend could be due to a transition between quasar-mode feedback generally observed more in higher-$z$ systems, to radio-mode feedback that is often characteristic of more low-redshift systems on average \citep[e.g.][]{2005MNRAS.363L..91C,2013MNRAS.432..530R,2017MNRAS.468.1398S}. Our current sample is not suited for weighing in on this trend, but we will investigate in a future paper whether there is any redshift evolution of $\epsilon_{\rm cool}$ in a larger, more complete, and unbiased Sunyaev-Zel'dovich (SZ)-selected sample (Calzadilla et al. in prep). 




\section{Discussion}\label{sec:discussion}

\subsection{A Gradual Saturation of AGN Feedback?}

The steeper-than-unity relation between SFR and ICM cooling rate presented in \autoref{subsec:SFRs} implies that multiphase condensation gradually becomes more prevalent as the maximal cooling rate increases. In this section, we attempt to explain only why the conversion from hot ($10^7$ K) to warm ($10^4$ K) phases becomes more efficient with cooling rate, and not whether the conversion from warm gas to stars in these systems is more efficient. In cool core clusters, the ICM density peaks sharply at the cluster center, where the condensing material accumulates within the BCG. This condensing material should eventually form cold molecular gas reservoirs that fuel star formation and accrete onto the central SMBHs. A large amount of the cooling in the most extreme cooling systems may be non-radiative (e.g. mixing/dust cooling). Regardless, despite clear evidence of feedback from the AGN (e.g. X-ray cavities), these outbursts
are unable to suppress cooling to the same degree as in systems with lower cooling rates. One way to understand why this is the case is to contrast the growth rates of the SMBHs versus that of the cluster halos they reside in. Using various empirical scaling relations, \citet{2018ApJ...858...45M} demonstrate that the accretion rate of SMBHs ($\dot{M}_{\bullet}$) scaled by the Eddington rate is related to the ICM cooling rate as $\dot{M}_{\bullet}/\dot{M}_{\rm Edd} \propto \dot{M}_{\rm cool}^{1.87}$, which is consistent with accretion rate data from \citet{2013MNRAS.432..530R}, and is supported by the tight positive scalings between the SMBH mass and hot halo properties \citep{2019ApJ...884..169G}. This correlation implies that more massive clusters, with very high cooling rates ($\mdot \gtrsim 10^3$ \msunperyear), should have central AGN accreting closer to the Eddington rate than in low-mass systems. 
However, while the left-hand side of the above relationship can ``saturate’’ as the SMBH growth rate is capped by the Eddington limit, the right-hand side has no analogous constraint. Galaxy cluster halos grow via mergers and accretion of smaller halos, which can relatively quickly increase the available amount of cooling material. In the most massive galaxy clusters, where AGN accretion approaches the Eddington rate, we might then expect disproportionately undermassive SMBHs and for ICM cooling to increasingly outpace the feedback at higher cooling rates, thus steepening the SFR-$\mdot$ relation. This is a testable hypothesis, though the direct measurement of SMBH masses and accretion rates in BCGs, especially those with quasar hosting systems, is difficult \citep[e.g.][]{2013ApJ...764..184M}.

Another consequence of the higher radiative efficiency of AGN in more strongly cooling halos is 
the resulting dominant mode of AGN feedback, which has an impact on how well the AGN energy can couple to the cooling ICM. As radiative efficiency $\dot{M}_{\bullet}/\dot{M}_{\rm Edd} \rightarrow 1$, a higher fraction of the AGN's accretion energy gets released in the form of radiation rather than in mechanical outflows via jets \citep{2005MNRAS.363L..91C,2013MNRAS.432..530R,2017MNRAS.468.1398S}. This transition from mechanical to radiative feedback is gradual and incomplete, meaning that it is not a sudden switch where radio jets turn off. Phoenix, H1821, and IRAS09104 are excellent examples of quasar-hosting systems that also have jet-inflated bubbles. Observationally, this radiative energy output begins to dominate gradually as the black hole's accretion rate approaches and exceeds a few percent of the Eddington rate, i.e. $\dot{M}_{\bullet} \gtrsim 0.1 \dot{M}_{\rm Edd}$, corresponding to a cooling rate of ${\sim}1000$ \msunperyear (see Fig. 8 in \citealt{2018ApJ...858...45M}). Above this accretion rate is where the quasar-hosting systems in our extreme cooling sample reside (see Fig. 12 in \citealt{2013MNRAS.432..530R}). We argue that it is no coincidence that the systems in our extreme cooling sample, particularly the quasar-hosting clusters, are also among the most highly efficient at converting hot gas into stars. An AGN in the radiative mode may allow more cooling to occur since the hot atmosphere it is embedded in is optically thin to radiation, making it less capable of coupling large amounts of its accretion energy to its surroundings and quenching cooling compared to a mechanical mode AGN. By contrast, radiatively inefficient, mechanical mode AGN can heat their surroundings via a number of simultaneous channels since their far-reaching jets can inflate bubbles which do work when expanding against their surroundings, as well as create shocks and sound waves, release cosmic rays, and mechanically increase the turbulence in the ICM \citep[e.g.][]{2001ApJ...554..261C,2017ApJ...847..106L,2002MNRAS.332..271R,2016ApJ...829...90Y,2001ApJ...549..832S,2014Natur.515...85Z,2017ApJ...845...91H,2019ApJ...871....6Y,2015MNRAS.451L..60G}. The fact that jets are `on' for a large fraction of the time  \citep[$\sim$60-70\%, e.g.][]{2006MNRAS.373..959D,2012MNRAS.427.3468B} is further evidence that specifically mechanical feedback is needed to regulate star formation and prevent runaway cooling. It would additionally suggest that radiatively efficient feedback is at least predominantly operating in, if not responsible for, those systems where cooling is overpowering heating.

The Eddington limit might play an even more critical role in systems with high-mass cores considering variations in the accretion flow onto the SMBH. If the \textit{average} accretion rate $\langle \dot{M}_{\bullet} \rangle \sim 0.1 \dot{M}_{\rm Edd}$, and accretion is clumpy and chaotic rather than steady, then any small clump of material that gets accreted will abruptly spike the accretion rate to the Eddington limit and suppress the most energetic outbursts via radiation pressure. In other words, at high average accretion rates, scatter can no longer be log normal because one side is truncated by the Eddington limit, while the other side is not.
Thus, the average effect of feedback may be reduced. Given this potential time-dependence, it is natural to ask whether the extreme cooling we see in our sample is a short lived phenomenon that occurs in most cool core clusters or if it is characteristic of only a small subset of cool core clusters. \citet{2021ApJ...910...60S} searched for Phoenix-like systems misidentified in the ROSAT survey \citep{1999A&A...349..389V} as X-ray bright point sources, and concluded that similar starbursting BCGs have an occurrence rate of $\lesssim 1\%$. Prior to that, \citet{2019ApJ...885...63M} used deep \textit{Chandra} data to show that the Phoenix cluster is the strongest example of a potential runaway cooling flow out of ${\sim}180$ cool core clusters ranging over 9 Gyr in time. Such an occurrence rate implies that if extreme cooling as seen in Phoenix is a common phenomenon, then it must only have a duration of roughly ${\sim}50$ Myr. This is consistent with simulations \citep[e.g.][]{2015ApJ...811..108P,2020MNRAS.495..594P} which show that episodes of high cooling rates and SFRs should last for ${\lesssim} 100$ Myr in any cool core cluster. However, the idea of cool-core cycles may be inconsistent with our findings of a slope steeper than unity in the SFR-$\mdot$ plot as well as the decreasing scatter with higher cooling rates shown in \autoref{fig:sfr_vs_mdot}. For instance, the lack of clusters with both $\mdot\,{\gtrsim}\,1000$ \msunperyear and SFR $\lesssim30$ \msunperyear, which should be relatively easy to detect and are missing even in the more complete sample compiled by  \citet{2018ApJ...858...45M}, tells us that our extreme cooling sample is probably not a sample of clusters caught at an opportune time (i.e. at a cooling extremum in the cool-core cycles described in \citealt{2015ApJ...811..108P,2020MNRAS.495..594P}). Still, the issue of how these extreme systems then stop cooling and quench star formation requires further investigation. It may be that with the more efficient conversion from cooling ICM to star formation that happens at high cooling rates ($\mdot\,{\gtrsim}\,1000$ \msunperyear), and thus closer to Eddington accretion rates ($\dot{M}_{\bullet} \gtrsim 0.1 \dot{M}_{\rm Edd}$), that the initially undersized SMBH eventually grows enough to increase the Eddington accretion rate itself, making the accretion fall into the sub-Eddington, mechanical feedback-dominated regime again. Indeed we do see ultra-massive black holes up to several $10^{10}$ \msun, which may push down the Eddington rates \citep[see][]{2019ApJ...884..169G}.
If we assume a CCA evolution, we often see spikes in accretion up to the Eddington regime lasting a few Myr, but given the chaotic nature of the feedback, the grown SMBH quickly self-regulates down to lower rates with a flickering time power spectrum described as in \citet{2017MNRAS.466..677G}.
More observational studies or simulations testing this hypothesis could be a promising path forward.


\begin{figure*}[ht]
\centering
\includegraphics[width=\textwidth]{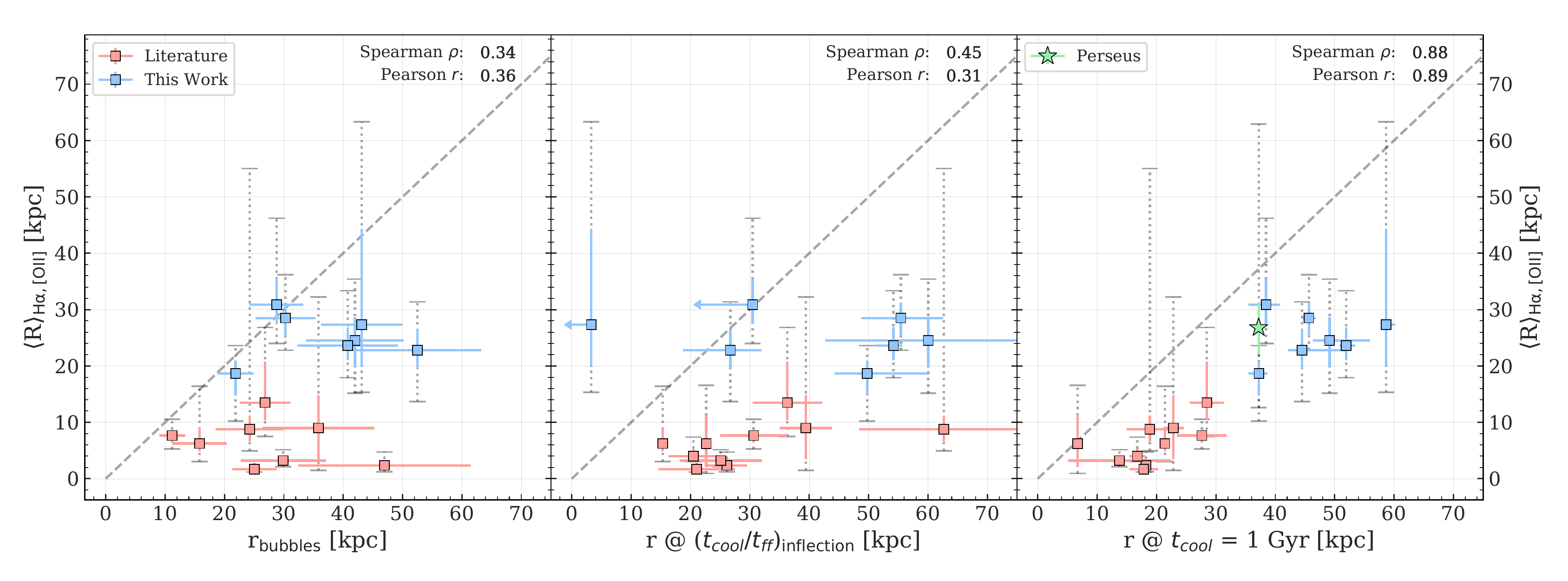}
\caption{Correlation plots between \textit{average} extent of multiphase gas ($\langle$R$\rangle_{\rm H\alpha, \OII}$ -- based on either \Ha (red points) or ${\rm \OII}$ emission (blue points), using maps from \citealt{2010ApJ...721.1262M} or this work, respectively) vs different proposed indicators of thermal instability from X-ray observations. In all panels, this average extent was calculated by measuring the maximum extent in ten equally-spaced angular bins, with the y-axis values representing the median over these ten measurements and the (colored) uncertainties representing the interquartile range (i.e. the 25\textsuperscript{th} and 75\textsuperscript{th} percentiles). These uncertainties are extended with vertical dotted lines to show the minima and maxima in these measurements. In all panels, a diagonal grey dashed line also represents a one-to-one relationship. \textit{\textbf{Left}}: $\langle$R$\rangle_{\rm H\alpha, \OII}$ vs extent of X-ray cavities (i.e. radio bubbles) from \citet{2008ApJ...687..173D}, using the distance to the cavity center as well as the leading edge distance for the uncertainties. 
\textit{\textbf{Middle}}: $\langle$R$\rangle_{\rm H\alpha, \OII}$ vs the radius where an inflection in the {\tcool/\tff} profile (modeled as two powerlaws) occurs. 
\textit{\textbf{Right}}: $\langle$R$\rangle_{\rm H\alpha, \OII}$ vs radius where {\tcool} = 1 Gyr. In \textit{middle} and \textit{right} panels, the x-axis values for the blue points come from modeling the ICM profiles in our extreme cooling sample, while values calculated using \citet{2017ApJ...851...66H} profiles are in red. Uncertainties for each are calculated via bootstrap resampling. In each of these panels, we calculate the Spearman-$\rho$ and Pearson-$r$ correlation coefficients between the two axes. 
The nearly total asymmetry to one side of the one-to-line in each of these panels demonstrates that all of these predictors of instability establish a volume within which multiphase cooling gas resides.
\label{fig:ROII_vs_everything} 
}
\end{figure*}

\subsection{Predicting the Onset of Thermal Instabilities}

The onset of thermal instabilities in the hot ICM is currently a contentious issue. 
Heating via mechanical feedback largely suppresses cooling in cool-core clusters, but \textit{some} cooling still happens, and moreover it is aided by this same self-sustaining feedback loop. Using the maps in \autoref{fig:OII_maps}, we have seen \textit{where} this cooling happens. Now, we can use these maps to compare the extent of the ${\rm \OII}$-emitting gas to recently-established metrics that attempt to explain the details of \textit{how} thermal instabilities develop in the presence of AGN feedback.
In this section, we will explore whether the measured extent of cool nebular gas in these maps correlates with X-ray derived radii that indicate cooling instability.
Extent measurements like these can be affected by projection along the line of sight, as well as observing depth, both of which lead to these measurements being lower limits of a true multiphase cooling threshold radius. To counteract these limitations, we add to the seven systems in our extreme cooling sample those of \citet{2010ApJ...721.1262M} for which we obtained \Ha maps.
We also utilize the \Ha map of Perseus  \citep[NGC1275:][]{2001AJ....122.2281C}, one of the closest, brightest, and most well-studied clusters. The network of \Ha filaments in Perseus may be representative of the range of extents and angles to our line-of-sight we could expect in other more distant systems. One caveat in Perseus (and possibly others) is that not all of the \Ha emission is associated with star formation \citep[e.g.][]{2010MNRAS.405..115C,2014MNRAS.444..336C}, with collisional heating being a potential ionization source instead \citep{2009MNRAS.392.1475F}. Even so, these \Ha and \OII maps of Perseus and others still trace where we see warm ${\sim}10^4$ K multiphase gas that has cooled out of an unstable hot atmosphere.

Rather than measure a single maximum extent of nebular gas in all of these systems, we measure the maximum extent in ten sectors, evenly spaced in azimuth and centered on the BCG position to find an \textit{average} extent, $\langle$R$\rangle_{\rm H\alpha, \OII}$. This more robust measurement further reduces the sensitivity that a single maximum extent has to small noise blobs at large radii. Additionally, it is more fair to compare an azimuthally-averaged extent to the azimuthally-averaged {\tcool} and {\tcool/\tff} profiles we will analyze in \autoref{subsubsec:R_OII_vs_profiles}.
These extent measurements are all listed in \autoref{tab:extent_data}, and plotted against different X-ray diagnostics of thermal instability in \autoref{fig:ROII_vs_everything}. In each of these panels, the shared y-axis measurements of $\langle$R$\rangle_{\rm H\alpha, \OII}$ have values and uncertainties (in solid colors) determined from the median and interquartile range (i.e. 25\textsuperscript{th} and 75\textsuperscript{th}) of the extent distributions, respectively. The larger, dashed, gray y-errorbars on each of these datapoints depicts the minimum and maximum extents across all azimuthal sectors for each cluster, which serves to encode the asymmetry of certain systems like Abell 1795, for instance, which are extended along mostly one direction or axis. In the following, we examine the relationship between these average extents and various indicators of ICM instability.

\subsubsection{$\langle R \rangle_{\rm H\alpha, \OII}$ vs Bubbles}
\label{subsubsec:R_OII_vs_bubbles}

One of the mechanisms by which AGN feedback may promote cooling is via radio bubbles inflated by the AGN. This ``stimulated feedback'' can be achieved by the wake transport of low-entropy warm gas by the buoyantly rising radio bubbles \citep[e.g.][]{2016ApJ...830...79M}. This uplifted warm gas has an increased infall time, which allows for in-situ production of cold ($10-100$ K) molecular gas. Joint observations of molecular gas with \textit{ALMA}, and of hot gas with \textit{Chandra} have revealed the molecular filaments appearing draped around X-ray cavities in a number of systems, with masses and kinematics consistent with uplift by the radio bubbles (e.g. Phoenix: \citealt{2017MNRAS.472.4024R}; Abell 1835: \citealt{2014ApJ...785...44M}; Abell 1664: \citealt{2019ApJ...875...65C}). One may expect that from our extreme cooling sample of clusters, we would see similarly intricate networks of filaments preferentially extended along the bubble/jet axis of AGN fueled by the strong cooling flows. However, as we showed in \autoref{fig:OII_maps}, the strongest cooling clusters have a diversity of extended \OII nebula morphologies, with some appearing chaotic rather than orderly as in an uplift scenario. For instance, comparing the Phoenix cluster to MACS1931, we see the \OII filaments extending mostly along the bubble axes in the former, and perpendicular in the latter.

To investigate this point further, we compare our extreme cooling sample to systems in the literature with measured X-ray cavity sizes and cluster-centric distances. Using the sample of \citet{2008ApJ...687..173D}, we collected the cavity size and location measurements (see \autoref{tab:extent_data}) for those clusters that have corresponding \Ha extent data from 
\citet{2010ApJ...721.1262M}. 
In a stimulated feedback scenario, multiphase gas may not extend beyond the maximum radius to which an AGN outburst can uplift low entropy gas.
However, in \autoref{fig:ROII_vs_everything} (left panel), we show that the cool gas radius ($\langle R \rangle_{\rm H\alpha, \OII}$ -- measured from the average extent of \OII or \Ha emission) has very weak correlation with the projected altitude of bubbles. The bubble distance is taken to be the average between the center of the X-ray cavity and its leading edge (i.e. cavity center plus cavity radius). The correlation strength is calculated via Pearson-$r$ and Spearman-$\rho$ coefficients, which are suited for assessment of exclusively linear or monotonically correlated data respectively. Both metrics are used to assess the presence and strength of a correlation non-parametrically, with an additional diagnostic of $\rho > r$ potentially indicating a non-linear relationship between two variables.
The weak correlation between bubbles and multiphase gas here ($\rho = 0.34$, $p_{\rm \rho}=0.22$; $r=0.36$, $p_{\rm r}=0.19$) suggests that while bubble uplift is likely a factor in promoting cooling in \textit{some} systems (e.g. in Perseus and Phoenix), it is not the whole story for most systems. 

This weak correlation should perhaps be unsurprising, as gas beneath cavities tends to be turbulent, and will follow the local pressure gradient. Gas filaments that condense out of uplifted low-entropy ICM material should eventually decouple from the bubble wake and fall back down to an altitude where the density contrast is minimized. There will always be such time evolution to the extents of the multiphase gas as well as the cavities that will wash out correlations, which are difficult to account for in observations (e.g. \citealt{2021ApJ...917L...7Q, 2021MNRAS.tmp.3347F}). To complicate matters further, some systems have multiple generations of X-ray cavities (e.g. Perseus: \citealt{2006MNRAS.366..417F}, Hydra: \citealt{2007ApJ...659.1153W}, NGC5813: \citealt{2011ApJ...726...86R}), sometimes even perpendicular to each other \citep[e.g. RBS797:][]{2021ApJ...923L..25U}, which make it difficult to connect a specific outburst to the extent of cooling seen at longer wavelengths. Conversely, the fact that the star-forming filaments in some systems extend beyond any detected X-ray cavities (e.g. Phoenix, A1795) does not necessarily rule out bubble uplift, but it is impossible to say without deeper data, and even that may not help when older bubbles are projected along the line of sight. 

Importantly, however, we can still learn from \autoref{fig:ROII_vs_everything} that filaments \textit{on average} always lie interior to the radius where we observe bubbles. Thus, the presence and locations of X-ray cavities is still a valuable metric for determining the radius within which the ICM becomes unstable. More measurements of the velocity structure of warm \Ha or \OII filaments, for instance with integral-field spectroscopy, or of turbulent motion in the ICM in the future (e.g. \citealt{2016Natur.535..117H,2016SPIE.9905E..2FB}) will help to further extend our understanding of bubble uplift.

\subsubsection{$\langle R \rangle_{\rm H\alpha, \OII}$ vs Cooling and Freefall Time Profiles}
\label{subsubsec:R_OII_vs_profiles}

In addition to the ``stimulated feedback'' model described above, other models predict that thermal instabilities can be produced by the cooling of the turbulent ICM when and where it becomes multiphase, i.e. where the specific entropy of the gas falls below some threshold value \citep[e.g.][]{2008ApJ...683L.107C}, resulting in ``precipitation'' of low angular momentum gas clouds onto the central SMBH. Precipitation models \citep[e.g.][]{2015Natur.519..203V,2012ApJ...746...94G} predict that these thermal instabilities develop within the transition radius between an outer baseline ICM entropy profile due to cosmological structure formation and an inner profile induced by chaotic cold accretion and feedback.

While models predict that thermally unstable cooling happens where {\tcool/\tff}$<10$, most {\tcool/\tff} profiles of cool core clusters, with the exception of Phoenix, do not fall significantly below this threshold, as seen in \citet{2017ApJ...851...66H}. Instead, \citet{2017ApJ...851...66H} argue that because mass profiles within the typical radii where {\tcool/\tff} approaches a minimum can be approximated by an isothermal sphere, and that entropy profiles within these typical radii are consistent with a single power law, then {\tcool/\tff} profiles should flatten out toward smaller radii. Furthermore, in most of the 33 \Ha emitting clusters studied by \citet{2017ApJ...851...66H}, the min({\tcool/\tff}) values were measured from an annulus just outside of a single inner core annulus with a noisy {\tcool/\tff} measurement, showing that these measurements are typically not well-resolved. For these reasons, we modeled each of the {\tcool/\tff} profiles for our extreme cooling sample (see \autoref{subsubsec:thermo_rprofiles}) as well as those from the \citet{2017ApJ...851...66H} clusters that have corresponding \Ha extent data from 
\citet{2010ApJ...721.1262M}
with the assumption of a floor value rather than a minimum. To do so, we fit two powerlaws to the {\tcool/\tff} profiles, fixing the slope of just the innermost powerlaw to zero and allowing both normalizations to vary. We performed each of these fits over 1000 bootstrap iterations to obtain a distribution of measurements of where the two powerlaws cross, and plot these inflection radii ($R(t_{\rm c}/t_{\rm ff})_{\rm inflection}$) against \Ha and \OII extents in \autoref{fig:ROII_vs_everything} (middle panel). We again find a weak correlation ($\rho=0.45$, $p_{\rm \rho}=0.07$ ; $r=0.31$, $p_{\rm r} = 0.22$), but note that the correlation strength is driven down largely by two upper limits on the inflection radius. These two upper limits include Phoenix, whose {\tcool/\tff} profile has no discernible floor or minimum in the \textit{Chandra} data, and H1821, whose bright point source in the X-ray observations prevented a resolved measurement of $R(t_{\rm c}/t_{\rm ff})_{\rm inflection}$. The fact that $\rho>r$ could be an indication of a slightly non-linear relationship between ($R(t_{\rm c}/t_{\rm ff})_{\rm inflection}$) and extent of multiphase gas. Just as in the bubble correlation plot discussed above, we see that most systems have an inflection point in their {\tcool/\tff} profile that surrounds the average volume within which we see cooling filaments. 

Some studies have found that the scatter of {\tff} values is significantly smaller than the range of {\tcool} values at either a fixed radius of 10 kpc \citep{2017ApJ...851...66H} or at altitudes where {\tcool/\tff} is at a minimum \citep{2016ApJ...830...79M}. These findings suggest that the local gravitational effects encoded in {\tff} do not add any predictive power for the onset of thermal instabilities over {\tcool} alone \citep{2017ApJ...851...66H}. Some difficulties also arise from calculating {\tff} profiles. In light of this, we also compare in \autoref{fig:ROII_vs_everything} (right panel) where the individual modeled {\tcool} profiles of \citet{2017ApJ...851...66H} first decrease past 1 Gyr versus the average multiphase cooling radius. \citet{2017ApJ...851...66H} show that
the deprojected {\tcool} profiles of clusters with no nebular emission do not go below this 1 Gyr threshold at a radius of 10 kpc, with the exception of Abell 2029.
We show that there is a very strong correlation between {\tcool} and the average extent of cooling ($\rho=0.88$, $p_{\rm \rho}=1.9\times 10^{-6}$; $r=0.89$, $p_{\rm r}= 8.1\times 10^{-7}$). The addition of our seven new extent measurements (plotted in blue) are especially helpful in establishing this correlation. Once more, we see that all of the average extents in this right panel of \autoref{fig:ROII_vs_everything} lie below the one-to-one line, indicating that we observe multiphase gas out to radii where the cooling time is $\lesssim$1\,Gyr. 
It could be argued that this cooling time threshold is somewhat arbitrary, as a much shorter threshold (e.g. 0.1 Gyr) would move all of the datapoints to the left, possibly above the one-to-one line in \autoref{fig:ROII_vs_everything}. This threshold should vary over the mass ranges of rich clusters down to poor groups, where central cooling times can be up to 10$\times$ shorter, highlighting the importance of some mixing timescale for normalization (e.g. {\tff} or $t_{\rm eddy}$).
Despite this, it is worth noting that the strong correlation between $\langle R \rangle_{\rm H\alpha, \OII}$ and $R(t_{\rm cool} = 1 {\rm ~Gyr})$ should persist and always arise in any scenario in which multiphase gas is supplied locally by the cooling of hot ambient gas, which nicely supports the notion that the presence of multiphase gas is linked to cooling of ambient gas, regardless of the mechanism. 

\begin{figure}[ht]
\centering
\includegraphics[width=\columnwidth]{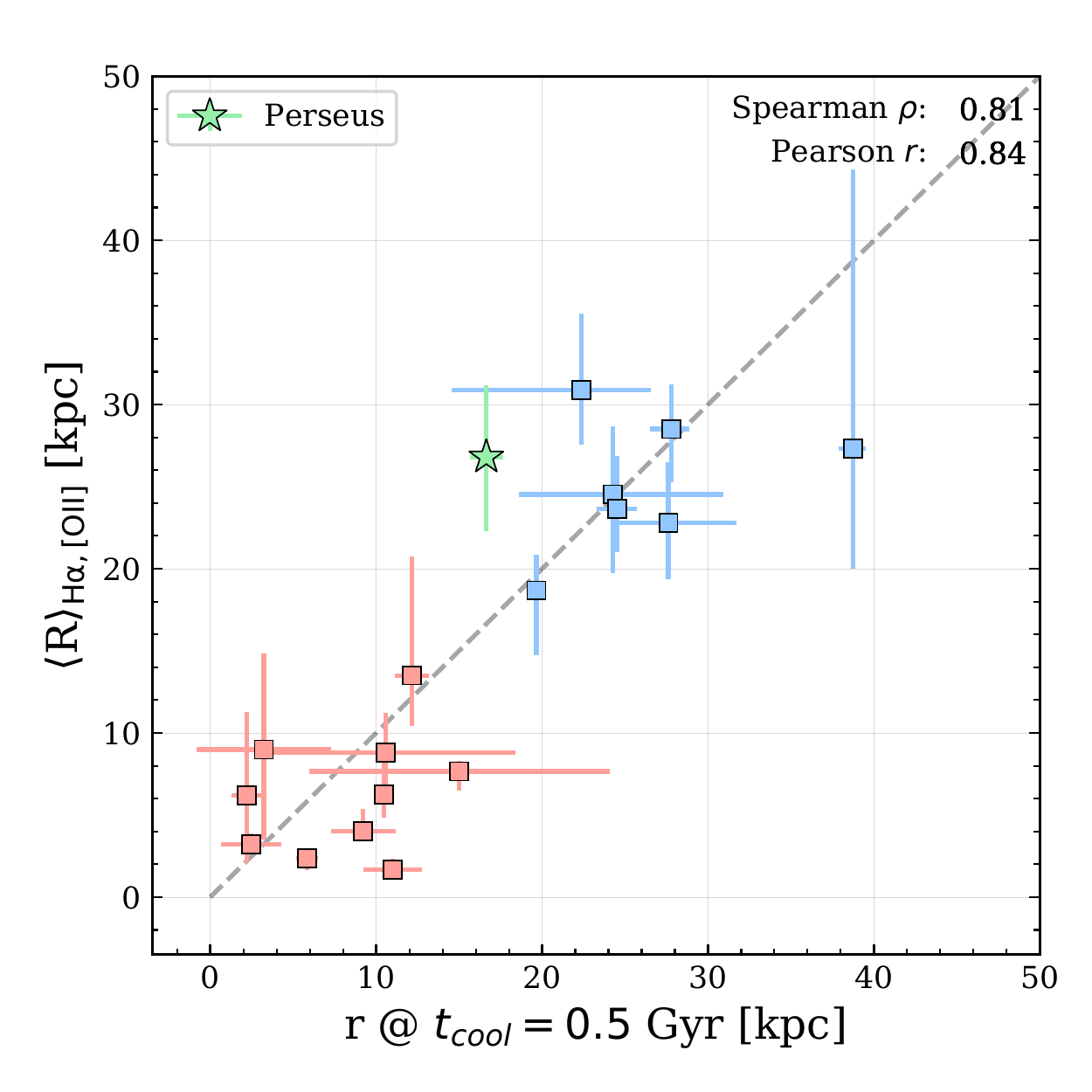}
\caption{Same as in \autoref{fig:ROII_vs_everything}, but focusing on the right panel correlation between $\langle R \rangle_{\rm H\alpha, \OII}$ and a different {\tcool} threshold. 
Extended y-errorbars showing the minimum and maximum extents of filaments have been omitted for clarity.
We find that measuring the radius at which {\tcool} reaches 0.5 Gyr (rather than 1 Gyr) has an approximately one-to-one correspondence with the extent of filaments.
\label{fig:ROII_vs_tcool_at_500Myr} }
\end{figure}

The strong correlation between $\langle R \rangle_{\rm H\alpha, \OII}$ and $R(t_{\rm cool} = 1 {\rm ~Gyr})$ in the right panel of \autoref{fig:ROII_vs_everything} is particularly interesting as it hints at the best estimate yet of where thermally unstable cooling ensues. By trying a range of cooling time thresholds below 1 Gyr, we find that measuring the radius at which {\tcool = 0.5} Gyr brings the datapoints closest to a one-to-one correspondence with the extent of optical filaments, as shown in \autoref{fig:ROII_vs_tcool_at_500Myr}. This tight relation suggests that, on average, when the ICM reaches a cooling time of 500 Myr or shorter, thermal instabilities will develop, leading to extended filaments of multiphase gas. 

More broadly, \autoref{fig:ROII_vs_everything} shows for the first time that multiphase gas on average resides  beneath the altitudes where we see X-ray cavities, the precipitation limit marked by a change in slope in {\tcool/\tff} profiles, and where the average cooling time is shorter than 1 Gyr. No matter which indicator one chooses, they all circumscribe the average volume within which the ICM is thermally unstable to multiphase cooling. The fact that the average extents of filaments do not lie on the one-to-one line with the altitudes defined by the various X-ray measurements is unsurprising as the snapshot in time at which we observe filaments at optical wavelengths could be drastically different from the onset of when the local cooling cascade in the ICM begins. In hydrodynamic simulations, cold gas is seen to quickly fall out of pressure equilibrium with hot gas (e.g. \citealt{2021ApJ...917L...7Q}). Turbulence could also play a large role not only in the altitude to which we observe multiphase gas, but also in the diverse morphology of \OII nebulae that we see in \autoref{fig:OII_maps} for instance. It could be that measuring the eddy timescale of the turbulent ICM \citep[e.g.][]{2018ApJ...854..167G} is more strongly correlated with the maximum extent of multiphase gas, but these measurements are difficult to make in practice, requiring filament kinematics. We may be able to more closely connect the hot (${>}10^7$ K) gas to the intermediate warm ($10^{5.5}$ K) gas, and better assess how ICM cooling flows fuel star formation, using future observations of coronal emission lines with instruments like MIRI aboard the \textit{James Webb Space Telescope}.







\section{Summary}\label{sec:conclusion}

In this work, we present new \textit{HST} observations of the \OII emission line nebulae in the strongest starbursting BCGs in the Universe. These new maps allow us to test the limits of AGN feedback in the presence of overwhelming cooling from the ICM. Together with archival \textit{Chandra} X-ray data, we can link the hot, X-ray emitting phase to this multiphase cooling gas to probe the development of thermal instabilities in the ICM. Our findings can be summarized as follows:

\begin{enumerate}
    \item \textit{HST} narrowband imaging of the \OII\doublet{3726,3729} emission-line doublet in the strongest known cooling clusters reveals massive filamentary nebulae of warm (${\sim}10^4$ K) gas extending $20-60$ kpc in altitude. These filaments have a wide range of morphologies, indicating a diversity or combination of creation mechanisms. In some cases, filaments are coincident with the rims of X-ray cavities, suggesting potential orderly uplift by buoyantly-rising radio bubbles, while in others the filaments are orientated perpendicular to the jet-bubble axis. Turbulence in the ICM may have a significant influence on the observed morphology of these nebulae.
    
    \item Our continuum-subtracted \OII maps have allowed us to secure more accurate integrated SFRs for our extreme cooling sample than previously available. With an average SFR of ${\sim} 150$ \msunperyear, these are among the strongest known starbursts in cluster cores. Combining these SFRs along with those of other systems from the literature, and comparing to maximal ICM cooling rates spanning 10 \msunperyear $< \mdot <$ 2000 \msunperyear, we find ${\rm \log(SFR)} = (1.67\pm0.17) \, {\rm \log(\mdot)} + (-3.25\pm0.38)$ with an intrinsic scatter of $0.39\pm0.09$ dex. This steeper-than-unity relationship means that the cooling of hot gas and the formation of young stars is most efficient in the strongest cool cores.
    
    \item This increasingly efficient conversion of hot ($\sim 10^7$ K) gas into warm star forming material implies a gradual decrease in the effectiveness of AGN feedback with higher ICM cooling rates. We propose that, as the cooling rate increases, the SMBH accretion rate will approach the Eddington limit, leading to an increasing fraction of the accretion energy released via radiation, rather than via the kinetic mode. The former is less effective at halting large-scale cooling, which would lead to an increase in the global SFR. Under this interpretation, it may not be a coincidence that the most efficiently cooling systems in our sample also host quasars.
    
    \item Using the average extent $\langle R \rangle$ of the multiphase gas measured from our \OII maps (along with \Ha measurements from the literature) as a proxy for \textit{where} the ICM has become thermally unstable, we compare to features in the ICM to assess how these instabilities develop. We show, for the first time, that multiphase gas on average resides beneath the altitudes where we see X-ray cavities, the precipitation limit marked by a change in slope in {\tcool/\tff} profiles, and where the average cooling time is shorter than 1 Gyr. No matter which indicator one chooses, they all circumscribe the average volume within which the ICM is thermally unstable to multiphase cooling.
    

    \item We find a strong correlation between $\langle R \rangle$ and the cooling radius of the hot ICM. Specifically, we find a one-to-one correlation between the average extent of the multiphase gas and the radius at which the ICM cooling time reaches 0.5 Gyr, which may be indicative of a universal condensation timescale in cluster cores.

\end{enumerate}

The new data presented here represent the sharpest view yet of the massive star forming regions in the strongest starbursting BCGs in the Universe. The unique environments provided by these systems allow us to test the limits of AGN feedback in the presence of overwhelming cooling. These systems could possibly mimic the environments of higher redshift cluster cores, as there is evidence that they are more likely to harbor central quasars as well as starbursts \citep[e.g.][]{2013MNRAS.431.1638H, 2022AJ....163..146S}, though possibly fueled somewhat differently \citep[e.g.][]{2016ApJ...817...86M}. Thus, this extreme cooling sample offers us a low-redshift window into higher-redshift phenomena, making these ideal candidates for future follow-up with observatories like the \textit{James Webb Space Telescope}.


\acknowledgments

This work is based on observations with the NASA/ESA Hubble Space Telescope obtained at the Space Telescope Science Institute, which is operated by the Associations of Universities for Research in Astronomy, Incorporated, under NASA contract NAS5-26555. These observations are associated with programs 15315, 15661, and 16001. 
MC acknowledges support from the NASA Headquarters under the Future Investigators in NASA Earth and Space Science and Technology (FINESST) award 20-Astro20-0037. 
MM and MC acknowledge financial support from programs HST-GO15315, HST-GO15661, and HST-GO-16001, which was provided by NASA through a grant from the Space Telescope Science Institute, which is operated by the Associations of Universities for Research in Astronomy, Incorporated, under NASA contract NAS5-26555. 
MM and MC acknowledge additional financial support for this work, provided by the National Aeronautics and Space Administration through Chandra Award Number GO0-21114A issued by the Chandra X-ray Center, which is operated by the Smithsonian Astrophysical Observatory for and on behalf of the National Aeronautics Space Administration under contract NAS8-03060. 
MG acknowledges partial support by NASA \text{Chandra} GO9-20114X and \textit{HST} GO-15890.020/023-A, and the \textit{BlackHoleWeather} program.

This work has made extensive use of the SAO/NASA Astrophysics Data System (\href{https://ui.adsabs.harvard.edu/}{ADS}) and the \href{https://arxiv.org/}{arXiv} preprint server.

%

\facilities{\textit{HST}(ACS), \textit{Chandra}(ACIS), CXO}


\software{astropy \citep{2013A&A...558A..33A},  
          matplotlib \citep{2007CSE.....9...90H}, 
          numpy \citep{2020Natur.585..357H}, 
          pandas \citep{2020zndo...3509134R}, 
          scipy \citep{2020NatMe..17..261V}, 
          seaborn \citep{2021JOSS....6.3021W}, 
          SExtractor \citep{1996A&AS..117..393B},
          CIAO \citep{2006SPIE.6270E..1VF}, 
          xspec \citep{1996ASPC..101...17A}
          }



\bibliography{sample63}{}
\bibliographystyle{aasjournal}

\appendix


\section{\Ha Homogenization and Contamination Correction}

\startlongtable
\begin{deluxetable*}{lrlrllll}
\centering
\tablenum{4}
\tablecaption{\Ha SFRs and $\dot{M}_{\rm cool}$ Data Used in \autoref{fig:sfr_vs_mdot}
\label{tab:Halpha_corrected}}
\tablewidth{0pt}
\tablehead{
\colhead{Name}  & \colhead{z} & \colhead{H$\alpha_{\rm lit}$} & \colhead{$L_{\rm H\alpha, homog.}$} & \colhead{$L_{\rm H\alpha, extinc.}$} & \colhead{$L_{\rm H\alpha, corr.}$} & \colhead{$\log_{10}{\rm SFR}_{H \alpha}$} & \colhead{$\log_{10}\dot{M}_{\rm cool}$} \\
\colhead{}  & \colhead{} & \colhead{} & \colhead{$10^{40}$erg s$^{-1}$} & \colhead{$10^{40}$erg s$^{-1}$} & \colhead{$10^{40}$erg s$^{-1}$} & \colhead{$\log_{10}$M$_{\odot}$ yr$^{-1}$} & \colhead{$\log_{10}$M$_{\odot}$ yr$^{-1}$}
}
\decimalcolnumbers
\startdata
2A 0335+096 & 0.035 & 8.3\tablenotemark{a} & 8.53  & 15$^{+7}_{-2}$ & 15$^{+7}_{-2}$ & 0.16$^{+0.16}_{-0.06}$ & 2.26$\pm$0.01 \\
3C295 & 0.464 & 2.2e-15\tablenotemark{h} & 134.32 & 240$^{+100}_{-30}$ & 240$^{+100}_{-30}$ & 1.4$^{+0.2}_{-0.1}$ & 2.98$\pm$0.04 \\
Abell 0085 & 0.056 & 1.6\tablenotemark{a} & 1.64  & 2.9$^{+1.3}_{-0.4}$ & 2.3$^{+1.3}_{-0.4}$ & -0.66$^{+0.19}_{-0.08}$ & 1.94$\pm$0.01 \\
Abell 0133 & 0.056 & 1.2\tablenotemark{a} & 1.23  & 2.2$^{+1.0}_{-0.3}$ & 1.8$^{+1.0}_{-0.3}$ & -0.77$^{+0.19}_{-0.08}$ & 1.79$\pm$0.01 \\
Abell 0478 & 0.088 & 23\tablenotemark{a} & 23.57 & 42$^{+18}_{-6}$ & 42$^{+18}_{-6}$ & 0.60$^{+0.16}_{-0.06}$ & 2.64$\pm$0.01 \\
Abell 0496 & 0.033 & 3.1\tablenotemark{a} & 3.18  & 5.7$^{+2.5}_{-0.8}$ & 5.3$^{+2.5}_{-0.8}$ & -0.29$^{+0.16}_{-0.07}$ & 1.75$\pm$0.03 \\
Abell 1204 & 0.171 & 59\tablenotemark{a} & 60.24 & 110$^{+50}_{-10}$ & 110$^{+50}_{-10}$ & 1.0$^{+0.2}_{-0.1}$ & 2.6$\pm$0.03 \\
Abell 1361 & 0.117 & 13.5\tablenotemark{d} & 6.89  & 12$^{+5}_{-2}$ & 12$^{+5}_{-2}$ & 0.067$^{+0.158}_{-0.062}$ & 1.66$\pm$0.19 \\
Abell 1413 & 0.143 & -2.38\tablenotemark{j} & -2.38 & $<$2.38 & $<$2.38 & $<$-0.42 & 1.74$\pm$0.12 \\
Abell 1650 & 0.084 & 0.03\tablenotemark{b} & 0.53  & 0.94$^{+0.41}_{-0.13}$ & 0.65$^{+0.41}_{-0.13}$ & -1.2$^{+0.2}_{-0.1}$ & 1.46$\pm$0.08 \\
Abell 1664 & 0.128 & 113.8\tablenotemark{d} & 58.06 & 100$^{+50}_{-10}$ & 100$^{+50}_{-10}$ & 1.00$^{+0.16}_{-0.06}$ & 2.21$\pm$0.04 \\
Abell 1689 & 0.184 & -0.28\tablenotemark{j} & -0.28 & $<$0.28 & $<$0.28 & $<$-1.50 & 2.31$\pm$0.1 \\
Abell 1795 & 0.063 & 1.13\tablenotemark{b} & 10.63 & 19$^{+8}_{-3}$ & 19$^{+8}_{-3}$ & 0.25$^{+0.16}_{-0.06}$ & 2.27$\pm$0.02 \\
Abell 1991 & 0.059 & 4\tablenotemark{a} & 4.10  & 7.3$^{+3.2}_{-1.0}$ & 7$^{+3.2}_{-1.0}$ & -0.17$^{+0.16}_{-0.06}$ & 1.58$\pm$0.04 \\
Abell 2052 & 0.035 & 1.8\tablenotemark{a} & 1.85  & 3.3$^{+1.4}_{-0.4}$ & 3.0$^{+1.4}_{-0.4}$ & -0.55$^{+0.17}_{-0.07}$ & 1.66$\pm$0.02 \\
Abell 2142 & 0.090 & 0.02\tablenotemark{b} & 0.40  & 0.72$^{+0.31}_{-0.10}$ & 0.39$^{+0.31}_{-0.10}$ & -1.4$^{+0.3}_{-0.1}$ & 1.73$\pm$0.07 \\
Abell 2199 & 0.030 & 2.7\tablenotemark{d} & 1.38  & 2.5$^{+1.1}_{-0.3}$ & 2.4$^{+1.1}_{-0.3}$ & -0.65$^{+0.16}_{-0.06}$ & 1.68$\pm$0.05 \\
Abell 2204 & 0.152 & 159.4\tablenotemark{d} & 81.33 & 150$^{+60}_{-20}$ & 150$^{+60}_{-20}$ & 1.1$^{+0.2}_{-0.1}$ & 2.7$\pm$0.01 \\
Abell 2244 & 0.097 & 0.333\tablenotemark{i} & 0.33  & 0.59$^{+0.26}_{-0.08}$ & 0.45$^{+0.26}_{-0.08}$ & -1.4$^{+0.2}_{-0.1}$ & 1.47$\pm$0.02 \\
Abell 2261 & 0.224 & 1.318\tablenotemark{i} & 1.32  & 2.4$^{+1.0}_{-0.3}$ & 2.0$^{+1.0}_{-0.3}$ & -0.72$^{+0.18}_{-0.07}$ & 2.1$\pm$0.1 \\
Abell 2390 & 0.230 & 109\tablenotemark{a} & 111.00 & 200$^{+90}_{-30}$ & 200$^{+90}_{-30}$ & 1.3$^{+0.2}_{-0.1}$ & 2.18$\pm$0.06 \\
Abell 2597 & 0.085 & 29.7\tablenotemark{j} & 53.84 & 96$^{+42}_{-13}$ & 96$^{+42}_{-13}$ & 0.96$^{+0.16}_{-0.06}$ & 2.49$\pm$0.05 \\
Abell 2626 & 0.057 & 1.1\tablenotemark{d} & 0.56  & 1.0$^{+0.4}_{-0.1}$ & 0.90$^{+0.44}_{-0.13}$ & -1.1$^{+0.2}_{-0.1}$ & 1.21$\pm$0.06 \\
Abell 3112 & 0.072 & 7.1\tablenotemark{a} & 7.28  & 13$^{+6}_{-2}$ & 12$^{+6}_{-2}$ & 0.074$^{+0.163}_{-0.065}$ & 1.93$\pm$0.05 \\
Abell 3581 & 0.022 & 2.4\tablenotemark{a} & 2.47  & 4.4$^{+1.9}_{-0.6}$ & 4.3$^{+1.9}_{-0.6}$ & -0.39$^{+0.16}_{-0.06}$ & 1.35$\pm$0.22 \\
Abell 4059 & 0.048 & 4.1\tablenotemark{a} & 4.21  & 7.5$^{+3.3}_{-1.0}$ & 7.0$^{+3.3}_{-1.0}$ & -0.18$^{+0.17}_{-0.07}$ & 1.09$\pm$0.06 \\
Cygnus A & 0.056 & 28.4\tablenotemark{j} & 21.32 & 38$^{+17}_{-5}$ & 38$^{+17}_{-5}$ & 0.56$^{+0.16}_{-0.06}$ & 2.15$\pm$0.01 \\
Hercules A & 0.154 & 29\tablenotemark{a} & 29.63 & 53$^{+23}_{-7}$ & 52$^{+23}_{-7}$ & 0.70$^{+0.16}_{-0.06}$ & 1.75$\pm$0.01 \\
Hydra A & 0.055 & 13\tablenotemark{a} & 13.34 & 24$^{+10}_{-3}$ & 24$^{+10}_{-3}$ & 0.35$^{+0.16}_{-0.06}$ & 2.04$\pm$0.02 \\
MKW3S & 0.045 & -14.7\tablenotemark{f} & 0.95  & 1.7$^{+0.7}_{-0.2}$ & 1.6$^{+0.7}_{-0.2}$ & -0.80$^{+0.16}_{-0.06}$ & 1.36$\pm$0.05 \\
MS 0735.6+7421 & 0.216 & 9\tablenotemark{g} & 93.40 & 170$^{+70}_{-20}$ & 170$^{+70}_{-20}$ & 1.2$^{+0.2}_{-0.1}$ & 2.42$\pm$0.08 \\
MS 1455.0+2232 & 0.259 & 29\tablenotemark{g} & 453.87 & 810$^{+350}_{-110}$ & 810$^{+350}_{-110}$ & 1.9$^{+0.2}_{-0.1}$ & 2.78$\pm$0.02 \\
NGC 4782 & 0.013 & 1.02\tablenotemark{c} & 0.78  & 1.4$^{+0.6}_{-0.2}$ & 1.1$^{+0.6}_{-0.2}$ & -0.99$^{+0.19}_{-0.08}$ & 0.23$\pm$0.03 \\
NGC 5044 & 0.009 & 0.54\tablenotemark{a} & 0.56  & 0.99$^{+0.43}_{-0.13}$ & 0.85$^{+0.43}_{-0.13}$ & -1.1$^{+0.2}_{-0.1}$ & 1.94$\pm$0.03 \\
PKS 0745-191 & 0.103 & 63\tablenotemark{a} & 64.51 & 120$^{+50}_{-20}$ & 110$^{+50}_{-20}$ & 1.0$^{+0.2}_{-0.1}$ & 2.89$\pm$0.01 \\
RXC J0352.9+1941 & 0.110 & 62\tablenotemark{a} & 63.47 & 110$^{+50}_{-20}$ & 110$^{+50}_{-20}$ & 1.0$^{+0.2}_{-0.1}$ & 2.3$\pm$0.03 \\
RXC J1459.4-1811 & 0.230 & 240\tablenotemark{a} & 244.39 & 440$^{+190}_{-60}$ & 440$^{+190}_{-60}$ & 1.6$^{+0.2}_{-0.1}$ & 2.48$\pm$0.04 \\
RXC J1524.2-3154 & 0.100 & 46\tablenotemark{a} & 47.11 & 84$^{+37}_{-11}$ & 84$^{+37}_{-11}$ & 0.90$^{+0.16}_{-0.06}$ & 2.23$\pm$0.01 \\
RXC J1558.3-1410 & 0.100 & 22\tablenotemark{a} & 22.53 & 40$^{+17}_{-5}$ & 40$^{+17}_{-5}$ & 0.58$^{+0.16}_{-0.06}$ & 2.1$\pm$0.03 \\
RXJ0439+0520 & 0.208 & 62\tablenotemark{a} & 63.20 & 110$^{+50}_{-10}$ & 110$^{+50}_{-10}$ & 1.0$^{+0.2}_{-0.1}$ & 2.39$\pm$0.23 \\
RXJ1504.1-0248 & 0.215 & 475.595\tablenotemark{i} & 475.60 & 850$^{+370}_{-110}$ & 850$^{+370}_{-110}$ & 1.9$^{+0.2}_{-0.1}$ & 3.29$\pm$0.08 \\
RXJ1539.5-8335 & 0.073 & 30\tablenotemark{a} & 30.76 & 55$^{+24}_{-7}$ & 55$^{+24}_{-7}$ & 0.72$^{+0.16}_{-0.06}$ & 2.19$\pm$0.05 \\
RXJ1720.1+2638 & 0.164 & 12.7\tablenotemark{d} & 6.48  & 12$^{+5}_{-2}$ & 11$^{+5}_{-2}$ & 0.038$^{+0.159}_{-0.063}$ & 2.63$\pm$0.03 \\
RXJ2129.6+0005 & 0.235 & 32\tablenotemark{a} & 32.58 & 58$^{+25}_{-8}$ & 57$^{+25}_{-8}$ & 0.74$^{+0.16}_{-0.06}$ & 2.36$\pm$0.33 \\
Sérsic 159-03 & 0.058 & 1.06\tablenotemark{b} & 8.53  & 15$^{+7}_{-2}$ & 15$^{+7}_{-2}$ & 0.15$^{+0.16}_{-0.06}$ & 2.37$\pm$0.02 \\
Zw 2701 & 0.210 & 8.7\tablenotemark{d} & 4.44  & 7.9$^{+3.4}_{-1.1}$ & 7.8$^{+3.4}_{-1.1}$ & -0.13$^{+0.16}_{-0.06}$ & 1.81$\pm$0.27 \\
Zw 3146 & 0.290 & 704.7\tablenotemark{d} & 359.55 & 640$^{+280}_{-90}$ & 640$^{+280}_{-90}$ & 1.8$^{+0.2}_{-0.1}$ & 2.87$\pm$0.11 \\
Perseus & 0.018 & & & & & 1.85$\pm$0.28\tablenotemark{k} & 2.67$\pm$0.05\\
\enddata
\tablecomments{Column 1: system name. Column 2: redshift. Column 3: Literature \Ha measurements, with varying measurements quoted (e.g. \Ha flux vs luminosity) and differing cosmologies (values of $H_0$, $\Omega_{\Lambda}$, $\Omega_m$) before homogenization to a \Ha luminosity (i.e. column 4):
$^a$\citet{2016MNRAS.460.1758H}, 
$^b$\citet{2010ApJ...721.1262M}, 
$^c$\citet{2018MNRAS.481.4472L}, 
$^d$\citet{1999MNRAS.306..857C}, 
$^e$\citet{2016AA...588A..68G}, 
$^f$\citet{2009AA...495.1033B}, 
$^g$\citet{1992ApJ...385...49D}, 
$^h$\citet{1989ApJ...338...48H}, 
$^i$\citet{2010MNRAS.401..433W}, 
$^j$\citet{2009ApJS..182...12C},
$^k$\citet{2015MNRAS.450.2564M}. Column 5: \Ha luminosity after intrinsic and Galactic extinction correction. Column 6: \Ha luminosity after correcting for evolved stellar population contribution (planetary nebulae, supernova remnants, AGB and HB stars, etc.). Column 7: \Ha star formation rate. Column 8: maximal (i.e. ``classical'') ICM cooling rate, from \citet{2018ApJ...858...45M}.
}
\end{deluxetable*}
\(\)

\clearpage

\section{Extent Data}

\startlongtable
\begin{deluxetable*}{lcccccc}
\centering
\tablenum{5}
\tablecaption{Various Extent Data Used in \autoref{fig:ROII_vs_everything}
\label{tab:extent_data}}
\tablewidth{5pt}
\tablehead{
\colhead{Name} & \multicolumn2c{r$_{\rm bubble}$} & \colhead{r @ ({\tcool/\tff})$_{\rm inflection}$} & \colhead{r @ {\tcool} = 1 Gyr} & \multicolumn2c{$R_{\rm H\alpha, \OII}$ [kpc]}  \\
\colhead{} & \colhead{(center) [kpc]} & \colhead{(center+radius) [kpc]} & \colhead{[kpc]} & \colhead{[kpc]} & \colhead{(median)} & \colhead{(min., max.)} 
}
\decimalcolnumbers
\startdata
Phoenix & 36.3 & 50.0 & $<$3.3\tablenotemark{a} & 59$^{+2}_{-1}$ & 27$^{+17}_{-7}$ & (15, 63) \\
H1821 & 24.2 & 33.3 & $<$30\tablenotemark{a} & 38$^{+2}_{-3}$ & 31$^{+5}_{-3}$ & (24, 46) \\
IRAS09104 & 41.7 & 63.3 & 27$^{+5}_{-8}$ & 45$^{+7}_{-2}$ & 23$^{+4}_{-3}$ & (14, 31) \\
Abell1835 & 18.8 & 25.0 & 50$^{+10}_{-6}$ & 37$^{+1}_{-2}$ & 19$^{+2}_{-4}$ & (10, 24) \\
MACS1931 & 25.3 & 35.4 & 55$^{+7}_{-7}$ & 46$\pm$1 & 29$\pm$3 & (23, 36) \\
RXJ1532 & 33.8 & 50.2 & 60$^{+16}_{-17}$ & 49$^{+7}_{-3}$ & 25$^{+4}_{-5}$ & (15, 35) \\
RBS797 & 32.3 & 49.2 & 54$^{+1}_{-3}$ & 52$\pm$1 & 24$\pm$3 & (18, 33) \\
\hline
A85   & 21.3 & 28.8 & 21$\pm$6 & 17$\pm$3 & 1.7$^{+0.6}_{-0.4}$ & (1.2, 3.2) \\
A133  & 32.4 & 61.5 & 26$\pm$3 & 18$\pm$1 & 2.4$^{+0.2}_{-0.7}$ & (1.2, 4.7) \\
A478  & 9.0  & 13.4 & 30$\pm$6 & 27$\pm$4 & 7.7$^{+0.7}_{-1.2}$ & (5.3, 11) \\
A496  &       &       & 21$\pm$4 & 17$\pm$2 & 4.0$^{+1.3}_{-0.6}$ & (3.0, 7.4) \\
A1644 &       &       & 23$\pm$3 & 7$\pm$1 & 6.2$^{+5.1}_{-4.1}$ & (0.9, 17) \\
A1795 & 18.5 & 30.0 & 64$\pm$13 & 20$\pm$5 & 8.8$^{+2.4}_{-2.2}$ & (4.9, 55) \\
A2052 & 11.2 & 20.4 & 15$\pm$3 & 21$\pm$1 & 6.3$^{+2.8}_{-1.4}$ & (3, 16) \\
A2597 & 22.6 & 31.1 & 36$\pm$6 & 28$\pm$3 & 14$^{+7}_{-3}$ & (8, 27) \\
A4059 & 22.7 & 37.1 & 26$\pm$8 & 13$\pm$9 & 3.2$^{+0.7}_{-0.5}$ & (2.1, 5.1) \\
Sérsic 159-03 & 26.4 & 45.3 &       &       & 9.0$^{+5.8}_{-5.5}$ & (1.5, 32) \\
Perseus\tablenotemark{b} &       &       &       & 37$\pm$1 & 27$^{+4}_{-5}$ & (13, 63) \\
\enddata
\tablecomments{Column 1: system name. Columns 2 and 3:
X-ray cavity/bubble data measured from our archival Chandra X-ray images (systems above horizontal rule), or taken from \citet{2008ApJ...687..173D}. Distance is measured to the bubble center (in column 2) as well as to the leading edge (center+radius; column 3) of the bubble. Column 4: inflection point in the {\tcool/\tff} profile, modeled as a double powerlaw with a floor (i.e. inner power law slope of zero). Column 5: radius where cooling time falls below 1 Gyr. Column 6: median and interquartile range (25\textsuperscript{th} and 75\textsuperscript{th} percentiles) of the distribution of optical filament extents measured from our \OII maps (i.e. systems above horizontal rule; see \autoref{fig:OII_maps}) or the \Ha measurements of
\citet{2010ApJ...721.1262M}. Column 7: minimum and maximum optical filament extents.
}
\tablenotetext{a}{The innermost radial bin from the X-ray data is quoted as an upper limit since the {\tcool/\tff} profile is consistent with a single power law, with no resolved inflection point in the profile.}
\tablenotetext{b}{Perseus \Ha data quoted from \citep{2001AJ....122.2281C,2003MNRAS.344L..48F}, and {\tcool} data from \citet{2006MNRAS.373..959D}. No bubble data are quoted for this system due to the multiple X-ray cavity pairs seen in Perseus, making the association between a particular cavity and the filaments difficult.}
\end{deluxetable*}
\(\)

\clearpage

\end{document}